\documentclass[a4paper]{article}
\pdfoutput=1
\usepackage{graphicx,wrapfig}
\usepackage{lipsum}
\usepackage{amsmath,amssymb,mathrsfs}
\usepackage{hyperref}
\usepackage{url}
\usepackage{mathtools}
\usepackage{changepage}
\usepackage{multirow}
\usepackage{wrapfig}
\usepackage{subfig}
\usepackage{xcolor}
\usepackage{epsfig,enumerate,amsmath,amsfonts,amssymb,amsthm,mathrsfs,ifpdf}
\usepackage{indentfirst,relsize}
\usepackage{setspace,graphicx}
\usepackage{latexsym}
\usepackage[all]{xy}
\usepackage[usenames,dvipsnames]{pstricks}
\usepackage{pst-grad} 
\usepackage{pst-plot} 
\usepackage[margin = 2.50cm]{geometry}
\usepackage[T1]{fontenc}
\usepackage[utf8]{inputenc} 

\usepackage{graphicx,booktabs,array,tabularx}
\usepackage{tikz}
\usetikzlibrary{calc,arrows.meta,positioning}
\usepackage[linesnumbered,ruled,vlined]{algorithm2e}
\usepackage{enumitem,authblk,microtype,needspace}

\usepackage[linesnumbered,ruled,vlined]{algorithm2e}

\usepackage{color,soul}

\usepackage[colorinlistoftodos]{todonotes}

\usepackage{authblk} 

\usepackage{orcidlink}

\usepackage{enumitem}

\usepackage{tikz}

\newcommand{\remove}[1]{}

\newtheorem{theorem}{Theorem}
\newtheorem{lemma}[theorem]{Lemma}
\newtheorem{proposition}[theorem]{Proposition}
\newtheorem{corollary}[theorem]{Corollary}

\newcounter{Case}[theorem]

\newcounter{Ca}[theorem]

\newcounter{NCa}[theorem]

\usepackage{xcolor}
\allowdisplaybreaks

\title{Proper Conflict-Free Coloring of Undirected Path, Chordal Bipartite, and Strongly Chordal Graphs}
\author[1]{Gautam K. Das\footnote{gkd@iitg.ac.in}}
\author[1]{Kamal Santra \orcidlink{0009-0006-5997-1452} \footnote{kamal.7.2013@gmail.com, kamal.santra@iitg.ac.in}}
\affil[1]{Department of Mathematics\\
	
	Indian Institute of Technology Guwahati\\
	
	Guwahati, 781039, Assam, India}
\date{}

\begin{document}

\maketitle
\begin{abstract}
	A proper conflict-free coloring is a proper coloring in which each
	nonisolated vertex has a neighbor whose color occurs exactly once in
	its open neighborhood. We prove that deciding whether the clique
	number of colors suffices is NP-complete on undirected path graphs,
	even with a supplied representation by paths in a star. One extra
	color suffices on every instance in the reduction. We also prove
	NP-completeness on chordal bipartite graphs for every fixed palette
	size $k\ge4$. For strongly chordal graphs, reverse simple elimination
	gives a constructive upper bound of one more than the clique number.
	We obtain exact polynomial-time algorithms under several conditions
	on the maximum cliques, including when they induce a connected
	subtree of a clique tree. For strictly chordal graphs we characterize
	the optimum and construct an optimal coloring in $O(n+m)$ time,
	where $n$ and $m$ count the vertices and edges. For strongly chordal graphs with clique-tree separators of size
	at most $s$, we give an exact algorithm in $n^{2^{O(s+1)}}$ time,
	with no bound on the clique number.
\end{abstract}



\section{Introduction}
\label{sec:intro}

In a proper coloring, adjacent vertices have different colors. A
proper conflict-free coloring also requires each nonisolated vertex
to have a neighbor whose color occurs exactly once in its open
neighborhood. Such a neighbor is called a witness. The difference
already appears on a three-vertex path. A proper two-coloring gives
both ends the same color, leaving the middle vertex without a
witness. A third color resolves this. The requirement is useful
when a vertex must identify a neighbor by its color alone, for
instance when colors represent communication channels.

We study how this local requirement affects the complexity of
coloring graphs with structured neighborhoods. For a graph $G$,
write $\omega(G)$ for its clique number and $\chi_{\mathrm{pcf}}(G)$
for its proper conflict-free chromatic number. Chordal graphs admit
an optimal ordinary coloring with $\omega(G)$ colors. With that
palette, each maximum clique uses every color. If a vertex belongs
to several maximum cliques, any witness for it must belong to their
intersection. Smaller maximal cliques can also affect the choice.
The colors they omit determine which witnesses in neighboring
cliques retain their uniqueness. On the bipartite side, ordinary
coloring needs at most two colors, which makes it natural to ask
whether the witness condition can make a fixed palette hard to
test even when induced cycles are restricted.

Fabrici et al. studied proper conflict-free
coloring alongside unique-maximum coloring~\cite{FabriciLuzarRindosovaSotak2023}.
Caro et al. investigated the relation to
ordinary coloring and determined the optimum on trees~\cite{CaroPetrusevskiSkrekovski2023}. Exact linear-time algorithms
are known for block and proper interval graphs~\cite{PradhanSharma2026,SharmaPaulPandey2026}. For fixed treewidth $t$,
Liu's bound of $2t+1$ colors and Courcelle's theorem yield an exact
algorithm. For each candidate palette, the coloring requirement
can be written in monadic second-order logic by specifying
independent color classes and a uniquely colored neighbor for
each nonisolated vertex~\cite{Courcelle1990,Liu2024}. This also gives
polynomial-time solvability on chordal graphs for each fixed $k$.
Graphs with $\omega(G)>k$ can be rejected, and the remaining graphs
have treewidth at most $k-1$. In earlier work, we proved hardness at the clique number on well-partitioned
chordal graphs and gave a clique-number-parameterized algorithm for
interval graphs~\cite{SantraStructured2026}. Ahn, Im, and Oum proved
that proper conflict-free $k$-colorability is NP-complete on
bipartite graphs for every fixed $k\ge3$~\cite{AhnImOum2025}.
Pradhan and Sharma established the same hardness on perfect
elimination bipartite graphs~\cite{PradhanSharma2026}.

\begin{figure}[!tbp]
	\centering
	\newcommand{\known}{\textit{Known}\enspace}
	\newcommand{\here}{\textit{Here}\enspace}
	\begin{tikzpicture}[
		x=1cm,y=1cm,
		class/.style={draw,rounded corners=2pt,align=center,font=\scriptsize,
			text width=3.44cm,inner xsep=3pt,inner ysep=3pt},
		wide/.style={text width=4.05cm},
		prior/.style={class,fill=black!5},
		result/.style={class,fill=black!13},
		branch/.style={-{Stealth[length=1.7mm,width=1.2mm]},semithick},
		elbow/.style={branch,rounded corners=2pt},
		panel/.style={anchor=west,font=\footnotesize\bfseries,inner sep=0pt}
		]
		\node[panel] at (0,0.20) {(a) Chordal graphs and subclasses};
		\node[prior,wide] (chordal) at (5.85,-0.65) {\textbf{Chordal}\\
			\known NP-complete at $\omega$~\cite{SantraStructured2026}};
		\node[prior] (wpc) at (1.88,-2.02) {\textbf{Well-partitioned}\\\textbf{chordal}\\
			\known NP-complete at $\omega$\\
			$\omega+1$ upper bound~\cite{SantraStructured2026}};
		\node[result] (path) at (5.85,-2.02) {\textbf{Undirected path}\\
			\here NP-complete at $\omega$\\
			Theorem~\ref{thm:path-hardness}};
		\node[result] (strong) at (11.80,-2.02) {\textbf{Strongly chordal}\\
			\here $\omega+1$ upper bound\\
			Theorem~\ref{thm:sc-additive}\\
			Exact complexity open};
		\node[prior] (split) at (1.88,-3.78) {\textbf{Split graphs}\\
			\known exact criteria~\cite{SharmaPaulPandey2026}};
		\node[prior] (interval) at (5.85,-3.78) {\textbf{Interval graphs}\\
			\known $\omega+1$ upper bound\\
			Exact FPT in $\omega$~\cite{SantraStructured2026}\\
			General complexity open};
		\node[result] (strict) at (9.82,-3.78) {\textbf{Strictly chordal}\\
			\here formula and linear time\\
			Thm.~\ref{thm:strictly-chordal-pcf}};
		\node[result] (maxima) at (13.79,-3.78) {\textbf{Maximum-clique}\\\textbf{conditions}\\
			\here polynomial time\\
			Section~\ref{sec:strong}};
		\node[prior] (proper) at (5.85,-5.37) {\textbf{Proper interval}\\
			\known exact linear time\\
			\cite{PradhanSharma2026,SharmaPaulPandey2026}};
		\node[prior] (block) at (9.82,-5.37) {\textbf{Block graphs}\\
			\known exact linear time~\cite{PradhanSharma2026}};
		\node[result] (sep) at (13.79,-5.53) {\textbf{Bounded clique-tree}\\\textbf{separators}\\
			\here exact algorithm\\
			Thm.~\ref{thm:strong-small-separators}};
		\node[prior] (forest) at (9.82,-6.64) {\textbf{Forests and trees}\\
			\known exact formula~\cite{CaroPetrusevskiSkrekovski2023}};
		\draw[branch] (chordal.south west) -- (wpc.north);
		\draw[branch] (chordal.south) -- (path.north);
		\draw[branch] (chordal.south east) -- (strong.north);
		\draw[branch] (wpc.south) -- (split.north);
		\draw[branch] (path.south) -- (interval.north);
		\draw[elbow] ($(strong.west)!0.6!(strong.south west)$)
		-- (7.84,-2.95) -- (interval.north east);
		\draw[branch] (strong.south west) -- (strict.north);
		\draw[branch] (strong.south east) -- (maxima.north);
		\draw[elbow] (strong.south) -- (11.80,-5.53) -- (sep.west);
		\draw[branch] (interval.south) -- (proper.north);
		\draw[branch] (strict.south) -- (block.north);
		\draw[branch] (block.south) -- (forest.north);
		
		\draw[black!40] (0,-7.32) -- (15.63,-7.32);
		\node[panel] at (0,-7.67) {(b) Bipartite graph classes};
		\node[prior] (bip) at (1.88,-8.64) {\textbf{Bipartite}\\
			\known NP-complete\\
			for fixed $k\ge3$~\cite{AhnImOum2025}};
		\node[prior] (peb) at (5.85,-8.64) {\textbf{Perfect elimination}\\\textbf{bipartite}\\
			\known NP-complete\\
			for fixed $k\ge3$~\cite{PradhanSharma2026}};
		\node[result] (cb) at (9.82,-8.64) {\textbf{Chordal bipartite}\\
			\here NP-complete\\
			for fixed $k\ge4$ (Thm.~\ref{thm:chordal-bipartite-hardness})};
		\node[prior] (convex) at (13.79,-8.64) {\textbf{Convex bipartite}\\
			\known logarithmic\\
			bound~\cite{SantraStructured2026}};
		\node[prior] (biconvex) at (13.79,-10.10) {\textbf{Biconvex}\\
			\known exact $O(n^{33})$\\
			Five colors (tight)~\cite{SantraStructured2026}};
		\node[prior] (bp) at (9.82,-10.10) {\textbf{Bipartite permutation}\\
			\known exact $O(n^{17})$~\cite{SantraStructured2026}\\
			Five colors (tight)~\cite{SharmaPaulPandey2026}};
		\node[prior] (chain) at (5.85,-10.10) {\textbf{Chain graphs}\\
			\known exact linear time~\cite{PradhanSharma2026}};
		\draw[branch] (bip.east) -- (peb.west);
		\draw[branch] (peb.east) -- (cb.west);
		\draw[branch] (cb.east) -- (convex.west);
		\draw[branch] (convex.south) -- (biconvex.north);
		\draw[branch] (biconvex.west) -- (bp.east);
		\draw[branch] (bp.west) -- (chain.east);
		
		\draw[black!40] (0,-10.80) -- (15.63,-10.80);
		\node[panel] at (0,-11.15) {(c) Bounded treewidth};
		\node[prior] (tw) at (1.88,-12.06) {\textbf{Treewidth $\le t$}\\
			\known polynomial time\\
			\cite{Courcelle1990,Liu2024}};
		\node[prior] (outer) at (5.85,-12.06) {\textbf{Outerplanar graphs}\\
			\known exact via treewidth\\
			\cite{Courcelle1990,Liu2024}};
		\node[prior] (cactus) at (9.82,-12.06) {\textbf{Cactus graphs}\\
			\known exact via treewidth\\
			\cite{Courcelle1990,Liu2024}};
		\node[class] (forestd) at (13.79,-12.06) {\textbf{Forests}\\
			See panel (a)};
		\draw[branch] (tw.east) -- (outer.west);
		\draw[branch] (outer.east) -- (cactus.west);
		\draw[branch] (cactus.east) -- (forestd.west);
	\end{tikzpicture}
	\caption{Selected graph-class inclusions and results for proper conflict-free coloring.
		Arrows point to subclasses. \emph{Known} indicates an earlier result or a consequence
		of one, and \emph{Here} indicates a contribution of this paper. The symbol $\omega$
		denotes the clique number.
		Split and block graphs are well-partitioned chordal~\cite{AhnJaffkeKwonLima2022}.
		In panel (c), $t\ge2$ is fixed.}
	\label{fig:class-hierarchy}
\end{figure}
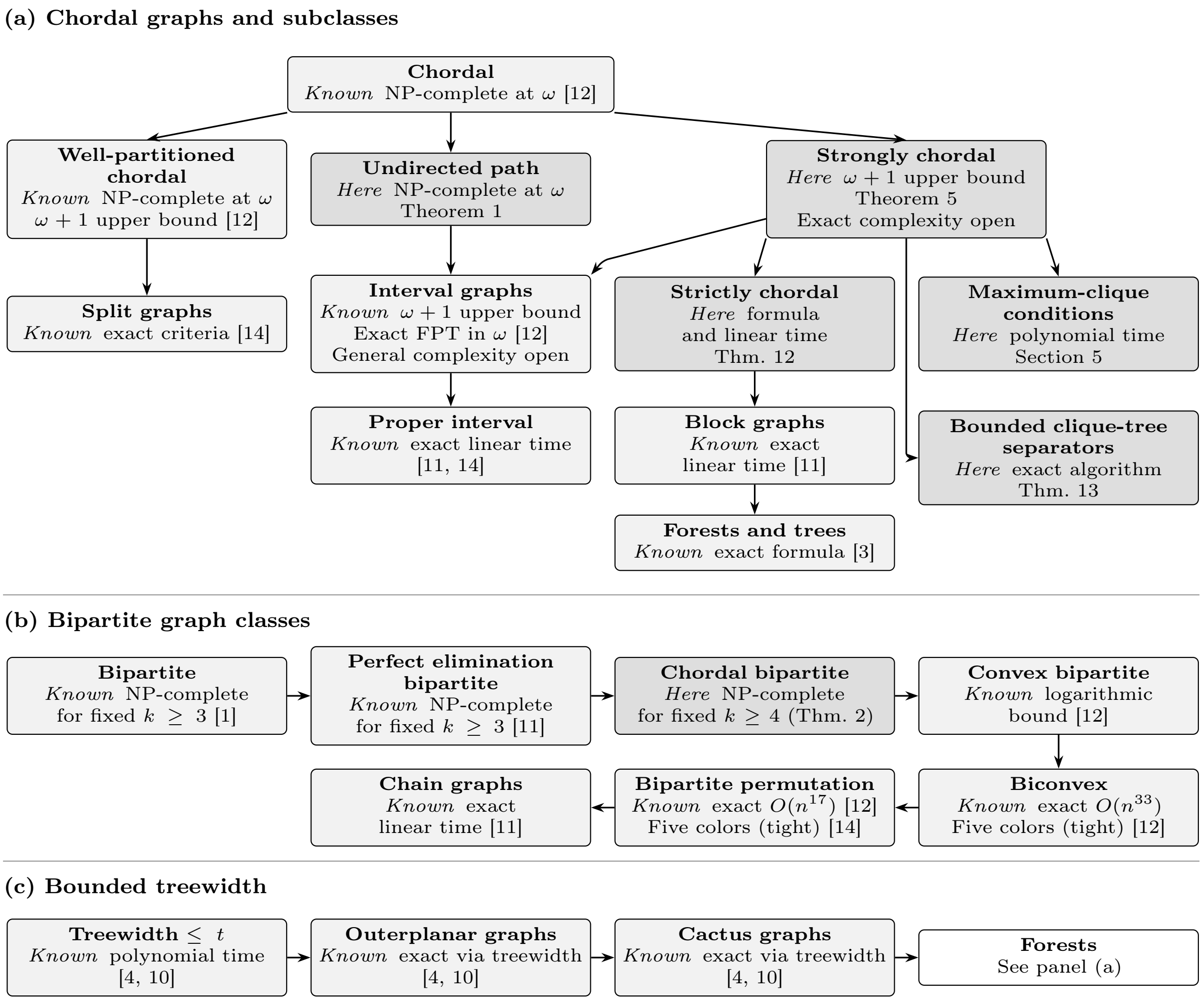

Figure~\ref{fig:class-hierarchy} places our results alongside the
relevant earlier ones.

\Needspace{7\baselineskip}
\medskip
\noindent\textbf{Our results}
\begin{enumerate}[leftmargin=*]
	\item \textbf{Undirected path graphs.}
	Deciding whether $\omega(G)$ colors suffice is NP-complete, even
	on connected graphs supplied with a representation by paths in a
	star. The instances in the reduction have optimum $\omega(G)$ or
	$\omega(G)+1$. The colors omitted by the leaf cliques encode
	perfect matchings in the variable gadgets
	(Theorem~\ref{thm:path-hardness}).
	
	\item \textbf{Chordal bipartite graphs.}
	Proper conflict-free $k$-colorability is NP-complete for every
	fixed $k\ge4$. The reductions impose color constraints through
	open neighborhoods and exclude induced cycles of length six
	or more (Theorem~\ref{thm:chordal-bipartite-hardness}).
	
	\item \textbf{Strongly chordal graphs.}
	Reverse simple elimination constructs a coloring with at most
	$\omega(G)+1$ colors in polynomial time
	(Theorem~\ref{thm:sc-additive}). We find an optimal coloring in
	polynomial time under several conditions on the maximum cliques.
	These include a nonempty intersection of all maximum cliques, a
	clique star with a maximum central clique, and maximum cliques
	that induce a connected subtree of a clique tree
	(Theorems~\ref{thm:sc-core} and~\ref{thm:sc-max-attachments} and
	Corollary~\ref{cor:sc-star}).
	
	\item \textbf{Strictly chordal graphs.}
	We construct an optimal coloring in $O(n+m)$ time. The optimum
	is $\omega(G)+1$ exactly when a vertex of the true-twin quotient
	representing a singleton class belongs to two blocks of maximum
	weight. In all other cases the optimum is $\omega(G)$
	(Theorem~\ref{thm:strictly-chordal-pcf}).
	
	\item \textbf{Bounded clique-tree separators.}
	For strongly chordal graphs with a supplied clique tree whose
	separators have size at most $s$, we find an optimal coloring in
	$n^{2^{O(s+1)}}$ time, with unrestricted clique number
	(Theorem~\ref{thm:strong-small-separators}). The algorithm groups
	colors by their truncated neighborhood counts on the separators.
\end{enumerate}

Section~\ref{sec:preliminaries} sets out the notation and graph
classes used in the proofs. Sections~\ref{sec:path-graphs}
and~\ref{sec:chordal-bipartite} establish the two hardness results.
Section~\ref{sec:strong} proves the additive-one bound and develops
the maximum-clique algorithms. Section~\ref{sec:strictly-chordal}
gives the linear-time algorithm for strictly chordal graphs, and
Section~\ref{sec:strong-small-separators} develops the algorithm for
bounded clique-tree separators. Section~\ref{sec:conclusion} discusses
the remaining complexity questions.

\section{Notation and preliminaries}
\label{sec:preliminaries}

All graphs are finite, simple, and undirected, except for an auxiliary
multigraph in Section~\ref{sec:path-graphs}. Write $N_G(v)$ and $N_G[v]$
for the open and closed neighborhoods of a vertex $v$. A
\emph{proper conflict-free $k$-coloring}, abbreviated PCF $k$-coloring,
is a proper coloring $c\colon V(G)\to [k]$ such that for every
$v$ with $N_G(v)\ne\varnothing$ there is a vertex $w\in N_G(v)$
whose color occurs once in $N_G(v)$. We call $w$ a \emph{witness}
for $v$. A palette of size $k$ may contain unused colors. We denote
the smallest feasible palette size by $\chi_{\mathrm{pcf}}(G)$ and
the clique number by $\omega(G)$. The empty graph has both parameters
zero. For a disjoint union, the PCF chromatic number is the maximum of
the values for its components.

A \emph{chordal graph} has no induced cycle of length at least four.
A maximal clique is inclusion-maximal. A maximum clique has
order $\omega(G)$. We use the standard chordal graph facts that
maximal cliques can be enumerated in polynomial time, there are at
most $|V(G)|$ of them, and an optimal proper coloring can be obtained
by coloring in reverse of a perfect elimination order. These facts
follow by deleting simplicial vertices, whose remaining neighbors
form a clique. The deleted vertex together with its remaining
neighbors contains any maximal clique for which it is the first
deleted vertex. A \emph{clique tree} has the maximal cliques as its
nodes, with the property that the cliques containing any given
vertex induce a connected subtree.

A $t$-\emph{sun}, for $t\ge3$, consists of a clique
$\{v_1,\ldots,v_t\}$ and an independent set
$\{u_1,\ldots,u_t\}$, with $N(u_i)=\{v_i,v_{i+1}\}$, where
indices are taken modulo $t$. A graph is \emph{strongly chordal}
if it is chordal and has no induced sun~\cite{Farber1983}.
A graph is an \emph{undirected path graph} if its vertices can be
represented by paths in a tree, with adjacency given by intersection
of the paths~\cite{Gavril1978}. Interval graphs are the special case
in which the host tree is a path. They are also strongly chordal.
Indeed, consider a clique of intervals meeting at a point. An
independent interval outside this clique that meets exactly two of
its members must lie wholly to one side of the point. On either
side, all pairs that can be met in this way coincide, since the
intervals are met in the order of their endpoints on that side.
A sun would require at least three distinct pairs. The reduction
in Section~\ref{sec:path-graphs} shows where these classes diverge.
Its instances can have an induced sun, even though they have
representations by paths in a star.

\section{Hardness on undirected path graphs}
\label{sec:path-graphs}

An undirected path graph is the intersection graph of a family of paths in a tree. We allow paths consisting of a single vertex, and distinct graph vertices may be represented by identical paths. The reduction below uses a star as the host tree. This restriction limits each path containing the center to at most two leaves, so a direct encoding in which a variable must occur at every leaf corresponding to one of its clauses would be unsuitable. Instead, we represent individual edges of an auxiliary multigraph by paths between pairs of leaves. The choice of colors omitted at a distinguished leaf encodes a truth assignment.

\begin{theorem}
	\label{thm:path-hardness}
	Deciding whether a graph \(G\) admits a proper conflict-free coloring with \(\omega(G)\) colors is NP-complete on connected undirected path graphs. Hardness holds for a supplied representation on a star, with \(\chi_{\mathrm{pcf}}(G)\in\{\omega(G),\omega(G)+1\}\).
\end{theorem}

\begin{proof}
	Given a coloring, one can check properness and count the occurrences of each color in each open neighborhood in polynomial time. The clique number can also be computed in polynomial time on this class. In a path representation, it equals the maximum number of paths containing any single host-tree vertex, by the Helly property of subtrees of a tree. Thus the problem belongs to NP.
	
	We reduce from \textsc{3-SAT}~\cite{Schaefer1978}. Write the variables as \(x_1,\ldots,x_r\) and the clauses as \(D_1,\ldots,D_m\). We may assume that \(r\ge1\) and that each clause contains between one and three literals on distinct variables. Indeed, a clause containing both signs of a variable can be dropped, and repeated occurrences of a literal can be deleted. An empty clause can be handled by outputting \(P_3\), which has clique number two and has no proper conflict-free two-coloring.
	
	\medskip
	\noindent\textbf{The auxiliary multigraph.}
	For each variable \(x_i\), create a private four-cycle with cyclic order \(z_{i,1},z_{i,2},z_{i,3},z_{i,4}\). Refer to its vertices as \emph{zero vertices}. Designate \(z_{i,1}z_{i,2}\) for the positive literal \(x_i\) and \(z_{i,2}z_{i,3}\) for the negative literal \(\neg x_i\). The two designated edges belong to opposite perfect matchings of the four-cycle. Replace every cycle edge by two parallel copies. The two copies will allow their corresponding graph vertices to witness one another without affecting which underlying cycle edges are selected.
	
	Add a vertex \(p\) adjacent to all \(4r\) zero vertices. For each clause \(D_j\), add a vertex \(q_j\) and the edge \(pq_j\). Also join \(q_j\) to the two endpoints of the designated edge of every literal of \(D_j\). These edges are simple. In particular, a clause on distinct variables connects \(q_j\) to exactly \(2|D_j|\) zero vertices. Denote the resulting multigraph by \(J\).
	
	\begin{figure}[t]
		\centering
		\begin{tikzpicture}[
			vertex/.style={circle,draw,fill=white,inner sep=1.5pt,minimum size=7mm,font=\small},
			note/.style={font=\small,align=center},
			edge/.style={line width=.55pt}]
			\node[vertex] (z1) at (-2.2,1.05) {$z_1$};
			\node[vertex] (z2) at (-.8,1.05) {$z_2$};
			\node[vertex] (z3) at (-.8,-.35) {$z_3$};
			\node[vertex] (z4) at (-2.2,-.35) {$z_4$};
			\node[vertex] (p) at (-3.65,.35) {$p$};
			\node[vertex] (q) at (.72,.35) {$q_j$};
			\foreach \a/\b in {z1/z2,z2/z3,z3/z4,z4/z1}{
				\draw[edge] (\a) to[bend left=10] (\b);
				\draw[edge] (\a) to[bend right=10] (\b);
			}
			\foreach \z in {z1,z2,z3,z4}
			\draw[edge,gray] (p) -- (\z);
			\draw[edge] (q) -- (z2);
			\draw[edge] (q) -- (z3);
			\draw[edge] (p) to[out=90,in=90,looseness=.8] (q);
			\node[note] at (-1.45,-1.26) {One variable cycle in \(J\)};
			\node[note] at (-1.45,-1.80) {Clause \(D_j\) uses \(\neg x_i\).};
			\begin{scope}[xshift=5mm]
				\node[vertex] (o) at (3.15,.35) {$o$};
				\node[vertex] (l1) at (2.2,1.35) {$i$};
				\node[vertex] (l2) at (4.1,1.35) {$j$};
				\node[vertex] (l3) at (2.2,-.65) {$h$};
				\node[vertex] (l4) at (4.1,-.65) {$\ell$};
				\draw[very thick] (l1)--(o)--(l2);
				\draw[edge,gray] (l3)--(o)--(l4);
				\node[note] at (3.15,-1.26) {Host star for \(G\)};
				\node[note] at (3.15,-1.80) {Bold path for \(ij\).};
			\end{scope}
		\end{tikzpicture}
		\caption{The two parts of the path-graph construction. On the left, the four vertices of one variable are joined by doubled cycle edges. The clause vertex is adjacent to the endpoints of its designated literal edge. Only the incidences involving this variable and this clause are shown. On the right, an edge of \(J\) gives a path through the center \(o\) of the host star.}
		\label{fig:path-reduction}
	\end{figure}
	
	\medskip
	\noindent\textbf{The path graph.}
	Create a clique \(C\) with a vertex for every edge copy of \(J\). Add \(2r+4\) padding vertices to \(C\), and put \(K=|C|\). For each vertex \(v\) of \(J\), let \(H_v\subseteq C\) be the set of vertices representing edge copies incident with \(v\). Define a nonnegative integer \(s_v\) by setting \(s_v=0\) for zero vertices, \(s_p=2r\), and \(s_{q_j}=|D_j|\). Add a clique \(L_v\) of size \(K-|H_v|-s_v\). Make \(L_v\) complete to \(H_v\) and anticomplete to \(C\setminus H_v\). There are no edges between distinct leaf cliques. The padding ensures every leaf clique is nonempty, since \(s_v\le\max\{2r,3\}\).
	
	Use a star with center \(o\) and one leaf for every vertex of \(J\). Represent a central vertex for an edge copy \(vw\) by the path \(v,o,w\), each padding vertex by \(\{o\}\), and each vertex in \(L_v\) by \(\{v\}\). Their intersections give exactly the stated adjacencies. Every \(H_v\) is nonempty, so \(G\) is connected. A clique that meets \(L_v\) lies in \(L_v\cup H_v\), which has size \(K-s_v\). As \(C\) has size \(K\), we have \(\omega(G)=K\). The number of edge copies and host vertices is \(O(r+m)\). Each leaf clique has at most \(K=O(r+m)\) vertices, so the construction is polynomial.
	
	\medskip
	\noindent\textbf{Omitted colors and witnesses.}
	In a proper \(K\)-coloring, identify a color with its unique representative in \(C\). The colors not used on \(L_v\) are exactly \(H_v\cup O_v\), where \(O_v\subseteq C\setminus H_v\) and \(|O_v|=s_v\). Conversely, for any such sets \(O_v\), coloring \(L_v\) bijectively with \(C\setminus(H_v\cup O_v)\) gives a proper coloring.
	
	Let \(e\in C\) represent an edge copy with endpoints \(v,w\). The vertex \(e\) sees every color except its own once in \(C\), and its only neighbors outside \(C\) lie in \(L_v\) and \(L_w\). Consequently, \(e\) has a witness if and only if
	\[
	\bigl((H_v\cup O_v)\cap(H_w\cup O_w)\bigr)\setminus\{e\}
	\ne\emptyset.
	\]
	If \(v\) is \(p\) or a clause vertex and \(z\) is a zero neighbor of \(v\), their unique edge \(vz\) gives \(H_v\cap H_z=\{vz\}\). Since \(O_z=\emptyset\), the edge vertex for \(vz\) has a witness exactly when \(O_v\cap H_z\ne\emptyset\).
	
	For each \(v\in\{p,q_1,\ldots,q_m\}\), the number of its zero neighbors is \(2s_v\). An omitted color belongs to the boundaries of at most two zero vertices, because it represents one edge copy or is a padding color. Therefore, meeting all \(2s_v\) required boundaries with \(s_v\) omitted colors forces every omitted color to meet exactly two required boundaries, and each required boundary to be met once. Equivalently, \(O_v\) consists of edge copies forming a perfect matching on the zero neighbors of \(v\). For \(p\), this means choosing one of the two underlying perfect matchings on each variable cycle. For \(q_j\), this means choosing one copy of each designated literal edge in \(D_j\). There are no other possible matching edges between its zero neighbors, because each variable occurs at most once in the clause.
	
	The unique token for \(pq_j\) is the only member of \(H_p\cap H_{q_j}\). Neither omission set meets the other endpoint's boundary, since both consist of edge copies joining two zero vertices. Thus the condition at \(pq_j\) reduces to \(O_p\cap O_{q_j}\ne\emptyset\). If a proper conflict-free \(K\)-coloring exists, read a truth assignment from the underlying matchings selected by \(O_p\). This intersection condition certifies a true literal in every clause.
	
	Conversely, fix a satisfying assignment. Use its matching on each variable cycle for \(O_p\), using a fixed copy of each edge. For each \(O_{q_j}\), select the fixed copy of every designated literal edge of that clause. These choices give witnesses for all edge vertices involving \(p\) or a clause vertex. The two copies of any cycle edge have identical closed neighborhoods, so they witness one another in a proper coloring. A padding vertex sees the clique \(C\) minus itself. A vertex of \(L_v\) sees the clique \((L_v\setminus\{x\})\cup H_v\), where \(x\) denotes that vertex. Both kinds of vertices therefore also have witnesses.
	
	Finally, \(K+1\) colors always suffice for the constructed graph. Give the vertices of \(C\) distinct colors and choose a padding vertex \(a\). In each \(L_v\), use distinct colors avoiding the colors on \(H_v\) and the color on \(a\). There are \(K-|H_v|\) available colors, enough for \(L_v\). The vertex \(a\) witnesses every other vertex of \(C\), and it is witnessed inside \(C\). The leaf vertices still see cliques. This proves the gap guarantee.
\end{proof}

The star representation in Theorem~\ref{thm:path-hardness} does not force strong chordality. In fact, the construction contains an induced \(3\)-sun. Take the three central vertices for \(pz,pz',zz'\), where \(zz'\) is a variable-cycle edge, and one vertex from each of \(L_p,L_z,L_{z'}\). The central vertices form a triangle, and the three independent leaf vertices see its three distinct pairs. The strongly chordal results therefore require a separate argument.
\section{Chordal bipartite graphs}
\label{sec:chordal-bipartite}

A bipartite graph is \emph{chordal bipartite} if it has no induced cycle of length at least six. This condition allows four-cycles. Ahn, Im, and Oum proved that \textsc{PCF \(k\)-Colorability} is NP-complete on bipartite graphs for every fixed \(k\ge 3\) \cite{AhnImOum2025}. The hardness results of Pradhan and Sharma for perfect elimination bipartite graphs do not give hardness for chordal bipartite graphs. Their reduction for three colors uses induced cycles of length at least six, and the construction for larger palettes retains an induced six-cycle from its subdivided palette clique \cite{PradhanSharma2026}. Here we give reductions whose output graphs are chordal bipartite.

\begin{theorem}
	\label{thm:chordal-bipartite-hardness}
	For every fixed integer \(k\ge 4\), \textnormal{\textsc{PCF \(k\)-Colorability}} is NP-complete on chordal bipartite graphs.
\end{theorem}

We reduce from two-colorability of a 3-uniform hypergraph. Equivalently, the input is a monotone Not-All-Equal formula whose clauses are triples of distinct variables, and the question is whether each triple can contain both Boolean values \cite{Schaefer1978}. For completeness, a two-element set \(\{a,b\}\) in the formulation with sets of size at most three can be replaced by the four triples \(\{a,b,c\},\{a,b,d\},\{a,b,e\},\{c,d,e\}\) with fresh \(c,d,e\). These triples can be split exactly when \(a\) and \(b\) have different values. A singleton or empty set gives an immediate negative instance. We use nontrivial inputs with at least one triple and at least three variables, since the remaining cases can be decided directly.

\subsection{Four colors}

Let \(A=\{a_1,\ldots,a_n\}\) contain one vertex for each variable. On the opposite part, introduce four vertices \(b_1,b_2,b_3,b_4\), called anchors, and two vertices \(g_j,h_j\) for each clause \(j\). Make \(A\) complete to the anchors and the clause vertices.

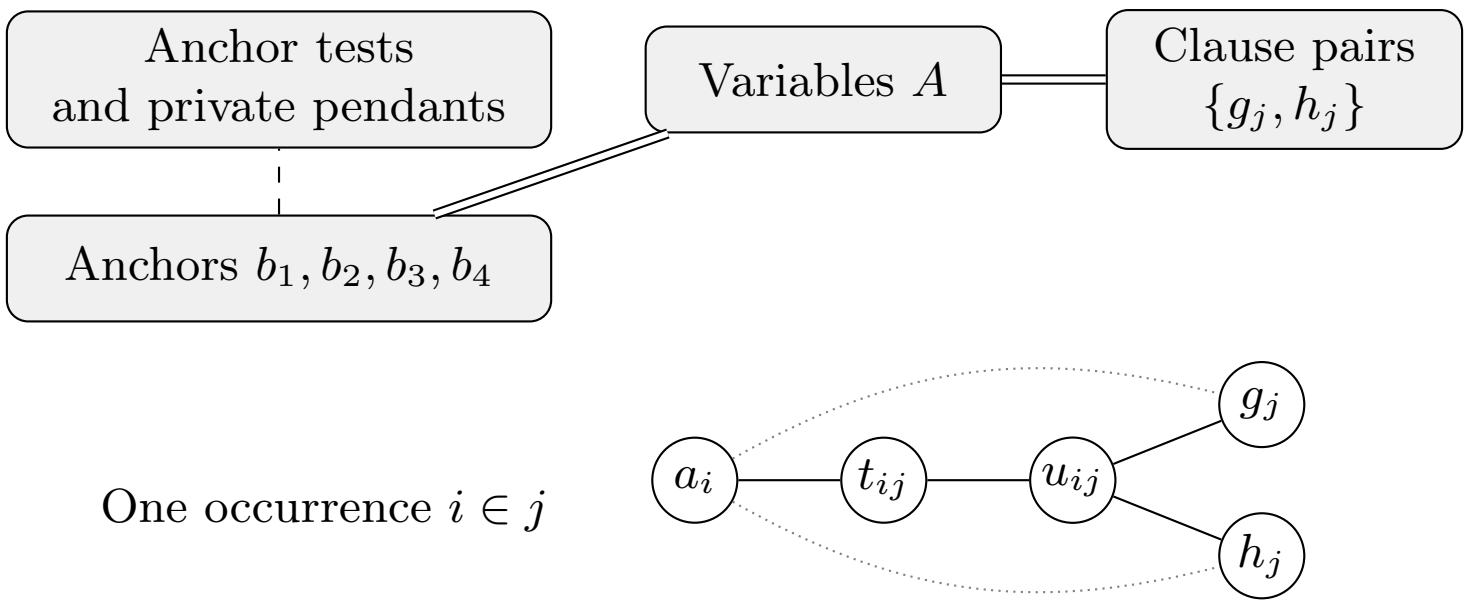
\begin{figure}[tbp]
	\centering
	\begin{tikzpicture}[x=1cm,y=1cm,
		vertex/.style={circle,draw,fill=white,inner sep=1pt,minimum size=16pt},
		group/.style={draw,rounded corners,fill=gray!12,minimum height=0.7cm},
		every node/.style={font=\small}]
		\node[group,minimum width=3.6cm] (B) at (-3.4,0.85) {Anchors \(b_1,b_2,b_3,b_4\)};
		\node[group,minimum width=3.6cm,align=center] (T) at (-3.4,2.1) {Anchor tests\\and private pendants};
		\node[group,minimum width=2.35cm] (A) at (0.2,2.1) {Variables \(A\)};
		\node[group,minimum width=2.35cm,align=center] (G) at (3.25,2.1) {Clause pairs\\\(\{g_j,h_j\}\)};
		\draw[dashed] (B)--(T);
		\draw[double, double distance=1.15pt] (A)--(G);
		\draw[double, double distance=1.15pt] (A)--(B);
		\node[vertex] (ai) at (-0.65,-0.55) {\(a_i\)};
		\node[vertex] (ti) at (0.6,-0.55) {\(t_{ij}\)};
		\node[vertex] (ui) at (1.85,-0.55) {\(u_{ij}\)};
		\node[vertex] (gj) at (3.1,-0.05) {\(g_j\)};
		\node[vertex] (hj) at (3.1,-1.05) {\(h_j\)};
		\draw (ai)--(ti)--(ui)--(gj) (ui)--(hj);
		\draw[gray,densely dotted] (ai) to[bend left=22] (gj);
		\draw[gray,densely dotted] (ai) to[bend right=22] (hj);
		\node[align=left] at (-3.1,-0.75) {One occurrence \(i\in j\)};
	\end{tikzpicture}
	\caption{The reduction for four colors. Double lines denote complete joins between the indicated sets. The dashed line indicates the anchor tests and pendant attachments described in the text. Dotted edges in the occurrence gadget are already present in the complete joins.}
	\label{fig:chordal-bipartite-gadget}
\end{figure}

On the part containing \(A\), add four test vertices with neighborhoods \(\{b_1,b_2\}\), \(\{b_2,b_3\}\), \(\{b_3,b_4\}\), and \(\{b_1,b_2,b_3,b_4\}\), respectively. Attach a private pendant vertex to each anchor. For each occurrence of variable \(i\) in clause \(j\), introduce two vertices \(t_{ij},u_{ij}\). Add the edges of the path \(a_i,t_{ij},u_{ij},g_j\) and the edge \(u_{ij}h_j\). The vertex \(t_{ij}\) lies on the part opposite \(A\), and \(u_{ij}\) lies on the same part as \(A\). In particular, \(t_{ij}\) has degree two, and \(N(u_{ij})=\{t_{ij},g_j,h_j\}\). There are no other edges. Figure~\ref{fig:chordal-bipartite-gadget} depicts the construction.

We first check the graph class. The following observation also applies to the reduction for larger palettes.

\begin{lemma}
	\label{lem:interval-tests}
	Let \(b_1,\ldots,b_p\) be distinct vertices in one part of a bipartite graph. Add test vertices, each adjacent to exactly the vertices of a nonempty interval in this order. The bipartite graph consisting only of the \(b_i\) and the test vertices is chordal bipartite. If the original graph is chordal bipartite and every original vertex adjacent to some \(b_i\) is adjacent to all the \(b_i\), then the resulting graph is chordal bipartite as well.
\end{lemma}

\begin{proof}
	In an induced cycle of length at least six in the graph of points and tests, each test on the cycle would have exactly two neighbors among the selected points. These points must be consecutive in the order restricted to the selected points. Contracting the two-edge steps through test vertices would therefore produce a cycle in the linear adjacency path of the selected points, which is impossible.
	
	Consider next an induced cycle using both a test vertex and a vertex outside the points and tests. The cycle must leave the test subgraph via a point and an original vertex adjacent to that point. Such an original vertex sees all the points, so at most two points can occur on the cycle, and both must be its cycle neighbors. There must in fact be two points, because entering and leaving the test subgraph requires two distinct points. The segment through the tests consists of a single test between these two points. Hence the cycle has length four.
\end{proof}

In the constructed graph, pendant vertices lie on no cycle. An induced cycle through \(t_{ij}\) must contain a segment \(a_i,t_{ij},u_{ij},y\) with \(y\in\{g_j,h_j\}\). The edge \(a_i y\) is a chord unless the cycle is a four-cycle. After ignoring all \(t_{ij}\), the pendant vertices, and the tests, every variable vertex is complete to the remaining opposite part. It cannot belong to an induced cycle of length at least six. Without the variables, only isolated anchors and disjoint complete bipartite clause graphs remain. Thus, the graph without the tests is chordal bipartite. Every neighbor of an anchor in this graph, before restoring the pendant vertices, is adjacent to all four anchors. Lemma~\ref{lem:interval-tests} permits us to restore the tests. Restoring the pendants introduces no cycle.

Suppose the graph admits a PCF coloring with at most four colors. A degree-two vertex has a witness exactly when its two neighbors have different colors. The three pair tests therefore force consecutive anchors to have different colors. With only two colors the sequence would alternate, and each color would occur twice at the full test. Hence the anchors use at least three colors. Since every variable vertex is adjacent to every anchor, the anchors use exactly three colors and every variable vertex uses the remaining color. Name this remaining color \(1\).

The vertex \(t_{ij}\) forces \(c(u_{ij})\ne 1\). The neighborhood of \(g_j\) includes all the variable vertices, so color \(1\) occurs there repeatedly. Its three occurrence neighbors avoid color \(1\) and the color of \(g_j\). Both available colors must occur among these three vertices, since otherwise no color occurs exactly once. Properness now forces \(g_j\) and \(h_j\) to use the same color, denoted by \(q_j\). At \(u_{ij}\), color \(q_j\) already occurs twice. Thus \(t_{ij}\) cannot use \(q_j\). The vertices \(t_{ij},u_{ij}\) use the two distinct colors outside \(\{1,q_j\}\).

The clause colors all agree. If all three colors other than \(1\) occur as clause colors, every variable vertex sees each of them at least twice on the clause pairs. If exactly two occur, let \(r\) be the third color other than \(1\). The only possible witness color at any variable is then \(r\). This color occurs on an anchor, so no \(t_{ij}\) can use \(r\). For any fixed clause \(j\), each \(t_{ij}\) must consequently use the same color, namely the color outside \(\{1,q_j,r\}\), and all three \(u_{ij}\) must use \(r\). This contradicts the condition at \(g_j\).

Rename the common clause color as \(2\). All occurrences now use signal colors \(3,4\), with opposite colors on \(t_{ij}\) and \(u_{ij}\). Each variable vertex sees both signals on the anchors and sees color \(2\) repeatedly on the clause pairs. If two of its occurrence vertices \(t_{ij}\) use different signals, neither signal can witness the variable. Thus the \(t_{ij}\) at a given variable all use the same signal. Read these signals as Boolean values and assign an arbitrary value to a variable with no occurrences. Each clause sees both signals on its \(u_{ij}\) vertices, so the assignment satisfies the formula.

Conversely, fix a satisfying assignment. Give the variables and the tests color \(1\), the clause pairs color \(2\), and the anchors the sequence \((2,3,2,4)\). For each occurrence, give \(t_{ij}\) the signal \(3\) or \(4\) assigned to its variable and give \(u_{ij}\) the other signal. Give each anchor pendant a color different from \(1\) and from its anchor's color. This is a proper coloring. Each variable sees the signal it does not use exactly once at an anchor. Each clause vertex sees a singleton among its occurrence neighbors because its triple uses both signals. The vertex \(t_{ij}\) sees two different colors, and its own color witnesses \(u_{ij}\). Each anchor is witnessed by its pendant, and the tests have witnesses by the anchor sequence. Every pendant sees its sole neighbor. This proves the equivalence for four colors.

\subsection{Larger fixed palettes}

We use a simple property of interval tests. If each nonempty interval of a sequence of \(p\) colored points contains a color occurring exactly once and the sequence uses at most \(r\) colors, then \(p\le 2^r-1\). To see this, choose a color occurring once in the full sequence. Each side of that occurrence uses at most \(r-1\) colors and satisfies the same property, so induction gives the bound. Conversely, the sequence on positions \(i=1,\ldots,2^{r-1}\) given by \(1+\nu_2(i)\), where \(\nu_2(i)\) is the exponent of two dividing \(i\), uses \(r\) colors and has a unique maximum color on every nonempty interval. The two largest ranks occur once each when \(r\ge 2\). All smaller ranks occur at least twice.

Fix \(k\ge5\) and start with the same formula. Introduce variable vertices \(A\) on one side and one clause vertex \(g_j\) per clause on the other side. Introduce the following three groups of points.
\begin{enumerate}
	\item A group \(B_0\) of \(2^{k-2}\) anchors on the clause side, with a test for every nonempty interval.
	\item A group \(A_0\) of four anchors on the variable side, with a test for every nonempty interval.
	\item A set \(F\) of support vertices on the clause side, partitioned into two groups of size \(2^{k-4}\) each. In each group, add a test for every nonempty interval in its own order.
\end{enumerate}
Give each anchor and each support vertex a private pendant neighbor. Make \(A\) complete to the clause vertices, \(B_0\), and \(F\). Make \(A_0\) complete to \(F\). For each occurrence of variable \(i\) in clause \(j\), add the path \(a_i,t_{ij},u_{ij},g_j\) and make \(u_{ij}\) complete to \(F\). These and the specified tests and pendants are all the edges.

We verify chordal bipartiteness before analyzing the colors. Again, no induced cycle of length at least six uses \(t_{ij}\). The cycle would have to include \(a_i,t_{ij},u_{ij},y\) for \(y=g_j\) or \(y\in F\), and \(a_i y\) would be a chord. Ignore tests and pendant vertices for the moment. Once the \(t_{ij}\) are removed, \(A\) is complete to the opposite part. After deleting \(A\), the support vertices are complete to the remaining opposite part. Deleting the supports leaves disjoint stars centered at the clause vertices and isolated vertices. Thus the graph at this stage has no induced cycle of length at least six. The vertices of each tested point group have the same external neighbors before restoring the tests and pendants. Restoring one group's tests does not change this property for any other group. Lemma~\ref{lem:interval-tests} proves that all the tests can be restored, and pendant vertices do not affect chordal bipartiteness.

In any PCF \(k\)-coloring the tests on \(B_0\) force at least \(k-1\) colors there. Hence the variables all use one color, which we name \(1\), and \(B_0\) uses every other color. The tests on \(A_0\) force at least three colors there, so the supports use at most \(k-3\) colors. Each support group uses at least \(k-3\) colors by its tests. Thus the two groups use the same palette \(P\) of exactly \(k-3\) colors, and every color in \(P\) occurs at least twice in \(F\). Since \(F\) is complete to \(A\), color \(1\) is not in \(P\). Let \(\sigma,\tau\) be the two colors outside \(P\cup\{1\}\).

The degree-two test at \(t_{ij}\) gives \(c(u_{ij})\ne 1\), and properness at \(F\) gives \(c(u_{ij})\notin P\). Thus \(u_{ij}\) uses \(\sigma\) or \(\tau\). Since \(g_j\) sees color \(1\) repeatedly, its triple must contain both signals. Properness forces \(c(g_j)\in P\). At \(u_{ij}\), every support color is already repeated, and \(g_j\) contributes another support color. Its witness must therefore be \(t_{ij}\) in the other signal color. Finally, \(a_i\) sees every color except \(1\) on \(B_0\) and every support color repeatedly on \(F\). A signal can witness \(a_i\) only if it is absent from all its occurrence vertices \(t_{ij}\). Therefore the occurrences of a variable agree, and the clauses use both signals as required.

Given a satisfying assignment, give \(A\) color \(1\). Color \(B_0\) by the rank sequence above, using \(\sigma,\tau\) for its two highest ranks and \(P\) for the other ranks. Color \(A_0\) by the same rank sequence with colors \(1,\sigma,\tau\), and color each support group by a rank sequence on \(P\). Give the clause vertices an arbitrary common color of \(P\). The interval tests on \(B_0\) and on the supports can all use color \(1\). The tests on \(A_0\) can all use a color of \(P\). Give each \(t_{ij}\) its variable's signal, and give \(u_{ij}\) the opposite signal. Each variable is witnessed by the unused signal at \(B_0\), each clause is witnessed by a signal in its triple, and \(t_{ij},u_{ij}\) have witnesses by the same neighborhood counts used above. A pendant at \(B_0\) can use any color different from \(1\) and its anchor's color. A pendant at \(A_0\) can use any other color in \(\{1,\sigma,\tau\}\), since all the nonpendant neighbors of that anchor use colors of \(P\). A pendant at a support vertex can use another color of \(P\), which exists because \(k\ge5\), and that color is absent from its nonpendant neighborhood. In each case the pendant color is absent from the nonpendant neighbors of its anchor or support vertex, so the pendant witnesses that vertex. Every pendant has its sole neighbor as a witness. The interval tests have witnesses by construction. This gives a PCF \(k\)-coloring.

For fixed \(k\), the anchor and support families have constant size. Both reductions have polynomial size, and a proposed coloring can be checked in polynomial time. This completes the proof of Theorem~\ref{thm:chordal-bipartite-hardness}.

\begin{proposition}
	\label{prop:chordal-bipartite-small-palettes}
	A bipartite graph admits a PCF two-coloring if and only if every nonisolated vertex has degree one. A nonempty chordal bipartite graph of minimum degree at least two has no PCF three-coloring.
\end{proposition}

\begin{proof}
	In a proper two-coloring every neighborhood is monochromatic. This gives the first claim. For the second, orient the edges of a proposed PCF three-coloring cyclically by colors, from \(1\) to \(2\), from \(2\) to \(3\), and from \(3\) to \(1\). Any directed cycle has length divisible by three, and bipartiteness forces its length to be even. A shortest directed cycle is induced, since any chord and the appropriate directed subpath would give a shorter directed cycle. Chordal bipartiteness therefore implies that the orientation is acyclic. However, at a vertex of degree at least two, the PCF condition requires neighbors of both other colors, so the vertex has positive indegree and positive outdegree. This is impossible in a finite acyclic digraph.
\end{proof}

Thus the hardness in this section concerns a fixed threshold strictly larger than the clique number. The complexity of PCF three-colorability on general chordal bipartite graphs is not determined by these results.
\section{Exact algorithms for subclasses of strongly chordal graphs}
\label{sec:strong}
We first show that a strongly chordal graph always has a PCF coloring
using at most one more color than its clique number. We then study when
that extra color is necessary. One exact test uses the intersection of the
maximum cliques. A second algorithm handles several maximum cliques
simultaneously, provided their vertices interact with the rest of the graph
through suitable clique attachments.

\subsection{Simple elimination and the additive-one bound}

A vertex \(x\) of a graph \(H\) is \emph{simple} if the closed
neighborhoods of the vertices in \(N_H[x]\) are linearly ordered by
inclusion. A simple vertex is simplicial. Indeed, for any two distinct
neighbors \(u,v\) of \(x\), the inclusion \(N_H[u]\subseteq N_H[v]\)
would imply \(uv\in E(H)\), and the reverse inclusion gives the same
conclusion. A \emph{simple elimination order} removes a simple vertex at
each step in the graph that remains. We use Farber's characterizations
of strongly chordal graphs by simple elimination and as chordal graphs
with no induced sun \cite{Farber1983}. We also use the fact that a
noncomplete strongly chordal graph has two nonadjacent simple vertices
\cite{DeCaria2012}. It follows that a simple elimination order can be
chosen to end with any prescribed clique \(R\). While vertices outside
\(R\) remain, remove any such vertex if the current graph is complete.
Otherwise at least one of the two nonadjacent simple vertices lies
outside \(R\).

The following observation will be used repeatedly. If \(v\ne w\) and
\(N_H[v]\subseteq N_H[w]\), then \(w\) is a witness for \(v\) in every
proper coloring of \(H\). In fact, \(v\) is adjacent to \(w\), and every
neighbor of \(v\) other than \(w\) is adjacent to \(w\). Properness
therefore makes the color of \(w\) unique in \(N_H(v)\).

\begin{theorem}\label{thm:sc-additive}
	Let \(G\) be a nonempty strongly chordal graph with clique number \(k\).
	Then \(k\le \chi_{\mathrm{pcf}}(G)\le k+1\).
	A coloring meeting the upper bound can be constructed in polynomial time.
\end{theorem}
\begin{proof}
	Write \(k=\omega(G)\), and color the vertices in reverse order of a simple
	elimination order. At each stage maintain a proper conflict-free coloring
	of the subgraph induced by the vertices already inserted. Let \(x\) be
	the next vertex. Its already colored neighbors form a clique of size at
	most \(k-1\). If this clique is empty, give \(x\) any color. Otherwise
	let \(z\) be a neighbor of \(x\) with largest closed neighborhood by
	inclusion in the graph after inserting \(x\). Such a neighbor exists
	by simplicity.
	
	For every neighbor \(v\) of \(x\) other than \(z\), we have
	\(N[v]\subseteq N[z]\) in the current graph. Thus \(z\) witnesses \(v\)
	once the coloring is proper. The vertex \(x\) also has a witness,
	since its neighborhood is a nonempty clique. If \(z\) previously had
	a witness, preserve it by avoiding its color on \(x\). If \(z\)
	was isolated before inserting \(x\), then \(x\) witnesses \(z\).
	In total we forbid the colors of the neighbors of \(x\) and at most
	one additional color. A palette of \(k+1\) colors suffices.
	
	The lower bound follows from properness. A simple elimination order
	can be found by testing closed-neighborhood inclusions and deleting a
	simple vertex at each step. With an adjacency matrix, testing every candidate directly at every
	stage takes $O(n^5)$ time on an $n$-vertex graph. The reverse
	steps use only neighborhood and color comparisons and fit within
	the same bound.
\end{proof}

\subsection{Maximum cliques with a common vertex}
\label{subsec:sc-core}

\begin{theorem}\label{thm:sc-core}
	Let \(G\) be a strongly chordal graph with clique number \(k\ge2\).
	Let \(I\) be the intersection of all maximum cliques of \(G\).
	If \(I=\{a\}\), then \(\chi_{\mathrm{pcf}}(G)=k+1\).
	If \(|I|\ge2\), then \(\chi_{\mathrm{pcf}}(G)=k\), and a PCF
	\(k\)-coloring can be found in polynomial time. In particular, a
	strongly chordal graph with a unique maximum clique admits a PCF
	\(k\)-coloring.
\end{theorem}
\begin{proof}
	Suppose first that \(I=\{a\}\). In any proper \(k\)-coloring,
	each maximum clique contains all \(k\) colors exactly once.
	If \(w\) witnesses \(a\), then \(w\) must belong to every maximum
	clique. Otherwise a different vertex of color \(c(w)\) in a maximum
	clique would also be a neighbor of \(a\). This is impossible since
	\(w\ne a\). Theorem~\ref{thm:sc-additive} now gives the equality.
	
	Suppose next that \(a,b\) are distinct vertices of \(I\). Use a simple
	elimination order ending with the edge \(ab\), and give \(a,b\) distinct
	colors. As we insert the other vertices in reverse, maintain that
	\(a\) witnesses every neighbor of \(a\) and \(b\) witnesses \(a\),
	as well as the PCF property. At the insertion of \(x\), forbid the
	colors on its already colored neighbors. We describe the extra
	protection that may be needed.
	
	If \(x\) is adjacent to \(a\), also forbid \(c(b)\).
	This preserves the witness of \(a\), and the color of \(a\) is
	unique in the clique neighborhood of \(x\). All previous neighbors
	of \(a\) retain their witness because \(x\) has a color different
	from \(c(a)\).
	
	If \(x\) is not adjacent to \(a\) but has a neighbor \(v\) adjacent
	to \(a\), let \(z\) be a neighbor of \(x\) with largest closed
	neighborhood in the current graph. Since \(N[v]\subseteq N[z]\),
	the vertex \(z\) is also adjacent to \(a\). Forbid \(c(a)\) on
	\(x\). This preserves the witnesses of all the affected neighbors
	of \(a\), including \(z\). Every other affected neighbor of \(x\)
	has \(z\) as a witness by closed-neighborhood inclusion.
	
	If \(x\) sees neither \(a\) nor any neighbor of \(a\), use the
	insertion rule of Theorem~\ref{thm:sc-additive}. This cannot
	disturb the invariant for \(a\) and its neighbors.
	
	In each case we forbid at most one extra color in addition to the
	colors on the clique neighborhood of \(x\). If this neighborhood
	has size \(k-1\), then \(x\) and its neighbors form a maximum
	clique. Both \(a\) and \(b\) belong to it, so \(x\) is in the
	first case and \(c(b)\) is already forbidden by properness.
	Otherwise the neighborhood has size at most \(k-2\).
	Thus a color always remains in the \(k\)-color palette, and the
	invariant holds throughout.
\end{proof}

For each \(k\ge2\), two copies of \(K_k\) sharing one vertex give a
strongly chordal graph with \(\chi_{\mathrm{pcf}}=k+1\). The
exclusive vertices of either copy are simple, so deleting them
one at a time gives a simple elimination order. The two maximum
cliques intersect in their common vertex, and
Theorem~\ref{thm:sc-core} applies.

\subsection{A polynomial witness-selection problem}
\label{subsec:sc-witness-selection}

Let \(U\) be a set of at least two labels and let
\(\mathcal H=(H_i)_{i\in J}\) be an indexed family of nonempty subsets
of \(U\). Each set \(H_i\) has a nonnegative integer budget \(d_i\).
We seek a map \(f\) with \(f(x)\in U\setminus\{x\}\) for each
\(x\in U\), subject to
\begin{equation}\label{eq:sc-budget}
	|f(H_i)\setminus H_i|\le d_i \qquad (i\in J).
\end{equation}
For a coloring application, a label will stand for a vertex whose
color is available as a witness. A budget bounds the number of
distinct witness colors that must be kept out of an attachment.

We call the incidence matrix of \(\mathcal H\) \emph{totally balanced}
if its bipartite incidence graph has no induced cycle of length at
least six. Equivalently, there is no square submatrix of order at
least three whose nonzero entries form one cycle. The maximal-clique
matrix of a strongly chordal graph is totally balanced
\cite{Farber1983}. Deleting rows or columns preserves this property.

\begin{lemma}\label{lem:sc-support-tree}
	If \(\mathcal H\) has a totally balanced incidence matrix, one can
	construct in polynomial time a tree \(T\) on \(U\) such that each
	\(H_i\) induces a connected subtree of \(T\).
\end{lemma}
\begin{proof}
	First we show that there is a label \(x\) for which the sets containing
	\(x\) are nested by inclusion. Make a split graph with a clique of
	row vertices indexed by \(J\) and an independent set of label
	vertices indexed by \(U\). A row vertex \(i\) is adjacent to \(x\)
	exactly when \(x\in H_i\). This graph is chordal. If it had an induced
	sun, all its clique vertices would be row vertices because an
	independent-set vertex is simplicial. Its independent rays would
	then be label vertices, and the incidences of the sun would give an
	induced cycle of length at least six in the incidence graph.
	Thus the split graph is strongly chordal and has a simple vertex.
	If a simple vertex is a label, its adjacent row vertices have
	nested closed neighborhoods, so their sets are nested.
	If a simple vertex is a row, its closed neighborhood contains
	every row vertex. In that case all row sets are nested, and any
	label works. This argument still applies after restricting the
	ground set to any subset of labels, since total balance is preserved.
	
	Successively remove a label \(x\) with this nesting property.
	If a set containing \(x\) has at least one other remaining label,
	join \(x\) to another label in the smallest such set, with all sets
	restricted to the labels still present. Otherwise join \(x\) to
	any label that remains after it. Do not add an edge for the final
	label. Each added edge joins a removed label to a later one,
	giving a tree. For each original \(H_i\), every member except the
	last removed member has its later neighbor in \(H_i\). Hence
	\(T[H_i]\) is connected. Direct tests of nesting give a
	polynomial construction. For $u=|U|$ and $m=|J|$, testing every
	candidate by pairwise set comparisons at every deletion takes
	$O(u^3(m+1)^2)$ time.
\end{proof}

\begin{lemma}\label{lem:sc-witness-greedy}
	If \(\mathcal H\) has a totally balanced incidence matrix, one can
	decide in polynomial time whether \eqref{eq:sc-budget} has a solution
	and construct a map when one exists.
\end{lemma}
\begin{proof}
	Take \(T\) from Lemma~\ref{lem:sc-support-tree}. We first show that
	we may require \(f(x)\) to be a tree neighbor of \(x\).
	Suppose some label uses \(y\) as its image at distance greater
	than one. In the component of \(T-y\) containing that label, let
	\(z\) be the neighbor of \(y\). Redirect to \(z\) all choices of
	\(y\) from that component, except the choice of \(z\) itself.
	
	Consider an affected boundary \(H_i\). If it contains \(y\),
	connectedness implies it also contains \(z\), so the change does
	not introduce an outside image. If it misses \(y\), it lies
	entirely in the component in question. When \(H_i\) contains \(z\),
	the new image is internal. When it also misses \(z\), every use
	of \(y\) in \(H_i\) has been replaced by \(z\), introducing at
	most one outside label in place of the previous one. In every
	case the left side of \eqref{eq:sc-budget} does not increase.
	The sum of the distances from labels to their images decreases.
	Repeating yields the desired map using only tree edges.
	
	Under this restriction, the left side of \eqref{eq:sc-budget}
	equals the number of labels in \(H_i\) choosing a neighbor outside
	\(H_i\). In fact, a tree vertex outside a connected subtree has at
	most one neighbor inside it, so different exits use different
	outside labels.
	
	Root \(T\), and process its vertices after their descendants.
	For each boundary its top is the vertex closest to the root.
	At \(x\), inspect the boundaries with top \(x\). Their contributions
	from descendants have already been determined. Reject if one of
	these exceeds its budget. If the contribution equals the budget,
	\(x\) must choose a neighbor inside that boundary. Intersect these
	requirements to obtain a set of feasible neighbors.
	
	If no boundary with top \(x\) is saturated and \(x\) has a parent,
	choose the parent. This cannot violate a boundary with top \(x\)
	and costs nothing in every pending boundary containing \(x\),
	since any such boundary also contains the parent. At the root
	there is no pending boundary containing it, so choose any
	feasible neighbor.
	
	It remains to justify the choice at a nonroot vertex with a
	saturated boundary \(R\) of top \(x\). Every feasible neighbor is
	a child in \(R\). Write \(p\) for the parent of \(x\).
	The memberships of feasible children in the pending boundaries
	containing \(x\) are nested. If two feasible children \(y,z\)
	had incomparable memberships, two pending boundaries would
	contain \(p,y\) but not \(z\), and \(p,z\) but not \(y\), respectively.
	The boundary \(R\) contains \(y,z\) but not \(p\). These three rows
	on columns \(p,y,z\) would form an induced incidence cycle of
	length six. Choose a feasible child whose membership is largest.
	It incurs no more cost in any pending boundary than another
	feasible child.
	
	To see that rejection is valid, start with any feasible map
	using tree neighbors. Inductively modify its choice at each
	processed vertex to match the algorithm. The budgets of all
	boundaries with top at that vertex remain satisfied. Every
	pending boundary is unchanged or weakly improved, and no
	previously settled boundary contains the vertex. Hence a feasible
	map can always follow the greedy decisions. This proves
	correctness and gives the constructive polynomial algorithm
	in Algorithm~\ref{alg:sc-witness-greedy}.
\end{proof}

\begin{algorithm}[t]
	\DontPrintSemicolon
	\KwIn{A totally balanced family \(\mathcal H\) on \(U\), and budgets \(d_i\).}
	\KwOut{A witness map satisfying \eqref{eq:sc-budget}, or a report of infeasibility.}
	Construct and root the tree \(T\) from Lemma~\ref{lem:sc-support-tree}\;
	\ForEach{\(x\) in an order in which descendants precede ancestors}{
		Set \(D_x\gets N_T(x)\)\;
		\ForEach{\(i\) for which \(x\) is the top of \(H_i\)}{
			Set \(a_i\gets |\{u\in H_i\setminus\{x\}: f(u)\notin H_i\}|\)\;
			\If{\(a_i>d_i\)}{\KwRet{infeasible}\;}
			\If{\(a_i=d_i\)}{\(D_x\gets D_x\cap H_i\)\;}
		}
		\If{\(D_x=\varnothing\)}{\KwRet{infeasible}\;}
		Set \(f(x)\) to a member of \(D_x\) contained in the largest number
		of pending boundaries containing \(x\)\;
	}
	\KwRet{\(f\)}\;
	\caption{Witness selection with boundary budgets.}
	\label{alg:sc-witness-greedy}
\end{algorithm}

\subsection{Witnessing a specified maximum clique}
\label{subsec:sc-designated}

\begin{theorem}\label{thm:sc-designated}
	Let \(G\) be strongly chordal and \(C\) a specified maximum clique
	of size \(k\ge2\). There is a polynomial-time algorithm to determine
	whether \(G\) has a proper \(k\)-coloring in which every vertex of
	\(C\) has a conflict-free witness. When the answer is yes, the
	algorithm constructs such a coloring.
\end{theorem}
\begin{proof}
	Identify the colors with their representatives in \(C\).
	In a proper \(k\)-coloring, any witness of a vertex \(x\in C\)
	must belong to \(C\setminus\{x\}\), because all colors other than
	\(c(x)\) already occur on its neighbors in \(C\). For each maximal
	clique \(Q\ne C\), use the boundary \(H_Q=Q\cap C\) and the
	budget \(d_Q=k-|Q|\). Empty boundaries impose no restriction.
	If a coloring supplies the witnesses \(f\) for \(C\), the vertices
	of \(Q\setminus C\) must avoid all colors corresponding to
	\(H_Q\cup f(H_Q)\). Since they form a clique, this requires
	\(|f(H_Q)\setminus H_Q|\le k-|Q|\).
	
	Conversely, suppose that \(f\) satisfies all these budgets.
	Root a clique tree of \(G\) at \(C\). If \(G\) is disconnected,
	its clique trees can be joined by edges with empty intersection.
	A vertex \(y\notin C\) first appears at a unique clique \(Q\)
	nearest to the root among the cliques containing it. Its neighbors
	in \(C\) are precisely \(H_Q\). Indeed, a clique containing \(y\)
	and a neighbor \(v\in C\) has \(Q\) on its path to the root.
	The running-intersection property then puts \(v\) in \(Q\).
	The converse follows since \(Q\) is a clique.
	
	Process the cliques from the root outward. For each new vertex
	at \(Q\), use the palette \(C\setminus(H_Q\cup f(H_Q))\).
	Along a root-to-leaf path the \(H_Q\) sets shrink, so these
	palettes grow. Thus already colored vertices in \(Q\setminus C\)
	use distinct colors in the current palette. The budget gives
	at least \(|Q\setminus C|\) colors in this palette, so we can
	assign distinct remaining colors to the new vertices. All earlier
	neighbors of a new vertex lie in \(Q\). Indeed, their clique
	subtrees meet the new vertex's subtree, and any such earlier
	subtree must pass through \(Q\) to meet it. Hence the coloring
	is proper. At the first appearance of any outside neighbor of
	\(x\in C\), we exclude the color of \(f(x)\). This makes \(f(x)\)
	a witness in the resulting graph.
	
	Finally, the incidence matrix of the sets \(H_Q\) is a
	submatrix of the maximal-clique matrix of \(G\). It is totally
	balanced by strong chordality. Lemma~\ref{lem:sc-witness-greedy}
	therefore decides the necessary and sufficient conditions and
	supplies \(f\) when possible.
\end{proof}

\begin{corollary}\label{cor:sc-star}
	An optimal PCF coloring can be computed in polynomial time on
	strongly chordal graphs with a maximum clique \(C\) such that every
	nonisolated vertex outside \(C\) has its closed neighborhood
	contained in that of another vertex. In particular, this applies
	to strongly chordal clique stars with a maximum central clique,
	with no restriction on the number of attachments or their budgets.
\end{corollary}
\begin{proof}
	The case \(\omega(G)=1\) is immediate.
	Test \(k=\omega(G)\) by Theorem~\ref{thm:sc-designated}. All other
	nonisolated vertices have witnesses by closed-neighborhood inclusion
	in any proper coloring. If the test rejects, use Theorem~\ref{thm:sc-additive}
	to construct an optimal coloring with \(k+1\) colors. In a clique
	star each leaf bag is complete to a subset of \(C\), and the leaf
	bags are pairwise anticomplete. Every vertex in a leaf bag is
	simplicial and has a witness in any proper coloring if it is nonisolated.
\end{proof}

\subsection{Interacting maximum cliques}
\label{subsec:sc-connected-max}

The next result also applies when the intersection of the maximum
cliques is empty. Write \(W\) for the union of all maximum cliques
of \(G\). For a component \(D\) of \(G-W\), let \(N_G(D)\cap W\)
denote the set of vertices of \(W\) with at least one neighbor in \(D\).

\begin{theorem}\label{thm:sc-max-attachments}
	Let \(G\) be strongly chordal and suppose that the following hold.
	\begin{enumerate}
		\item Every edge of \(G[W]\) belongs to a maximum clique of \(G\).
		\item For each component \(D\) of \(G-W\), the set \(N_G(D)\cap W\)
		is contained in one maximum clique of \(G\).
	\end{enumerate}
	Then an optimal PCF coloring of \(G\) can be computed in polynomial
	time. These hypotheses hold when the maximum cliques induce a
	connected subtree in some clique tree of \(G\).
\end{theorem}
\begin{proof}
	The cases \(k=\omega(G)\le1\) are immediate, so assume \(k\ge2\).
	As before, any witness \(f(x)\) of \(x\in W\) in a proper
	\(k\)-coloring belongs to every maximum clique containing \(x\).
	For each maximum clique \(Q\), impose a zero budget for the
	boundary \(Q\subseteq W\), forcing \(f(Q)\subseteq Q\).
	For each maximal clique \(Q\) containing a vertex outside \(W\),
	use boundary \(H_Q=Q\cap W\) and budget \(d_Q=k-|Q|\).
	Omit empty boundaries, which impose no constraint.
	
	These latter constraints are necessary. All vertices of \(Q\setminus W\)
	belong to one component \(D\) of \(G-W\), because they form a clique.
	Let \(C_D\) be a maximum clique containing \(N_G(D)\cap W\).
	The sets \(H_Q\) and \(f(H_Q)\) lie in \(C_D\), so their union
	has pairwise different colors. Each vertex in \(Q\setminus W\)
	must avoid those colors to preserve properness and the chosen
	witnesses. Hence \(|f(H_Q)\setminus H_Q|\le k-|Q|\).
	The incidence matrix of all these boundaries is again a submatrix
	of the maximal-clique matrix of \(G\). We can therefore use
	Lemma~\ref{lem:sc-witness-greedy} to select \(f\) or reject.
	
	Assume that we obtain \(f\). First color \(G[W]\) properly with
	\(k\) colors, which is possible by chordality. For every \(x\in W\)
	the chosen \(f(x)\) is already a witness among its neighbors in
	\(W\). To see this, let \(y\in N_G(x)\cap W\) be distinct from
	\(f(x)\). By condition 1, the edge \(xy\) lies in a maximum
	clique. This clique contains \(f(x)\) by its zero-budget
	constraint. Properness gives \(c(y)\ne c(f(x))\).
	
	Now process each component \(D\) independently. Choose a simple
	elimination order of \(G[D\cup C_D]\) ending at \(C_D\). Insert the
	vertices of \(D\) in reverse order while keeping the coloring of
	\(W\) fixed. Let \(x\in D\) be the next vertex, let \(A\) be its
	already colored neighbors, and let \(z\) be a member of \(A\)
	with largest closed neighborhood by inclusion in the current
	induced subgraph of \(G[D\cup C_D]\), if \(A\) is nonempty.
	All neighbors of \(D\) outside \(D\) lie in \(C_D\), so \(A\)
	is the full set of already colored neighbors of \(x\) in \(G\).
	It is a clique.
	
	If \(x\) has a neighbor in \(C_D\), then \(z\in C_D\).
	Otherwise \(z\in D\) and its closed neighborhood contains that
	of a neighbor of \(x\) in \(C_D\), so \(z\) is complete to the
	\(k\)-clique \(C_D\), a contradiction. In this case, every
	affected neighbor of \(x\) outside \(C_D\) is witnessed by \(z\)
	after the insertion. This follows from closed-neighborhood
	inclusion in the current graph on \(D\cup C_D\) and the fact
	that such a neighbor has no neighbors outside \(D\cup C_D\).
	
	We give \(x\) a color that avoids both the colors on \(A\) and
	the witness colors for \(A\cap C_D\). Let \(Q\) be a maximal
	clique containing \(A\cup\{x\}\). Since all of \(W\) is already
	colored, \(H_Q=Q\cap W=A\cap C_D\). Consequently, the number of
	forbidden colors is at most
	\(|Q|-1+|f(H_Q)\setminus H_Q|\le k-1\).
	This ensures an available color. It also preserves every required
	witness in \(C_D\). The new vertex \(x\) sees a clique, so it
	has a witness.
	
	If \(x\) has no neighbor in \(C_D\), use the insertion rule of
	Theorem~\ref{thm:sc-additive}. The vertex \(x\) is outside every
	maximum clique, so \(A\cup\{x\}\) has at most \(k-1\) vertices.
	There is room in the \(k\)-color palette to avoid at most one
	extra color in addition to the colors of \(A\). If \(A\) is empty,
	any color suffices. No witness in \(W\) is affected in this case.
	
	The construction gives a PCF \(k\)-coloring. If the budget test
	rejects, no such coloring exists, and Theorem~\ref{thm:sc-additive}
	constructs an optimal \((k+1)\)-coloring. The tests and all the
	constructions above are polynomial.
	
	Finally, suppose that the maximum cliques induce a connected
	subtree of a clique tree. Any branch outside this subtree attaches
	at a maximum clique \(C\). A vertex of \(W\) occurring in that
	branch must belong to \(C\) by the running-intersection property.
	In particular, any edge with both endpoints in \(W\) that occurs
	in a clique of the branch has both endpoints in \(C\). This proves
	condition 1. The clique subtree of a vertex outside \(W\) lies
	wholly in one branch, and two adjacent vertices outside \(W\)
	have intersecting clique subtrees. Hence a component of \(G-W\)
	lies in one branch and can meet \(W\) only through its attachment
	clique. This proves condition 2.
\end{proof}

\subsection{Why the single-clique tests do not suffice}
\label{subsec:sc-local-example}

Even for an interval graph, testing every maximum clique separately
by Theorem~\ref{thm:sc-designated} can give a positive answer at each
clique when the graph has no PCF coloring with the clique number of
colors. We give an explicit example.

Start with the seven intervals below.
\begin{center}
	\begin{tabular}{@{}cl@{\qquad}cl@{}}
		\toprule
		Vertex & Interval & Vertex & Interval \\
		\midrule
		$v$ & $[-1,17]$ & $u$ & $[12,31]$ \\
		$w$ & $[-1,13]$ & $z$ & $[16,31]$ \\
		$p,q$ & $[9,21]$ & $x$ & $[14,15]$ \\
		\bottomrule
	\end{tabular}
\end{center}
At each of $0,10,20,30$ add a group of short intervals, all of radius
$1/4$ about that point. Denote the groups by $A,B,C,D$, respectively,
with sizes $4,2,2,4$. The four maximum cliques are
\[
M_A=A\cup\{v,w\},\quad M_B=B\cup\{v,w,p,q\},\quad
M_C=C\cup\{u,z,p,q\},\quad M_D=D\cup\{u,z\}.
\]
The other maximal cliques are $\{v,w,p,q,u\}$,
$\{v,x,p,q,u\}$, and $\{v,z,p,q,u\}$. These lists follow by
sweeping the interval endpoints. In particular, the clique number is
six.

Here are feasible witness maps for each of the four single-clique
tests. In $M_A$, pair $v$ with $w$, and pair the four vertices of $A$
into two disjoint pairs. In $M_B$, pair $v$ with $w$ and $p$ with $q$,
and pair the two vertices of $B$. In $M_C$, pair $u$ with $z$ and $p$
with $q$, and pair the two vertices of $C$. In $M_D$, pair $u$ with
$z$ and pair the four vertices of $D$ into two disjoint pairs. Each
pair selects its two members as witnesses for one another.
The intersection of a chosen maximum clique with any other maximum
clique is a union of these pairs. Its zero budget is therefore met.
At any smaller maximal clique, at most one pair is split, so its
budget of one is also met.

Nevertheless, $v$ must use $w$ as a witness in a PCF six-coloring,
since $M_A\cap M_B=\{v,w\}$. Similarly $u$ must use $z$.
Since $z$ is a neighbor of $v$, the two witness colors $c(w)$ and
$c(z)$ must differ. The five vertices $v,u,p,q,x$ form a clique and
must avoid both colors. For $p,q,x$ this follows because they are
neighbors of both $v$ and $u$. For $v$ and $u$ it follows from
properness and from the witness condition at the other vertex.
The five-clique would then have only four available colors, a
contradiction. By Theorem~\ref{thm:sc-additive}, the graph has PCF
chromatic number seven. The example isolates a compatibility
constraint that is invisible to the single-clique tests.
\section{An exact algorithm for strictly chordal graphs}
\label{sec:strictly-chordal}

A block graph is a graph in which every block is a clique. We count
each bridge as a block of order two and each isolated vertex as a
block of order one. A strictly chordal graph is obtained from a
block graph by replacing each vertex with a nonempty clique and
joining two replacement cliques completely whenever the
corresponding vertices are adjacent~\cite{DumasEtAl2021}. In the
algorithm we use the quotient $B$ obtained by identifying vertices
with equal closed neighborhoods in $G$. For $x\in V(B)$, let $C_x$
be the true-twin class represented by $x$.

To see why $B$ is a block graph, consider any representation of
$G$ by clique replacements of a block graph. Each replacement
clique is contained in a single true-twin class. Choose one
representative of each true-twin class. They come from distinct
replacement cliques, so the graph they induce is isomorphic to an
induced subgraph of the original block graph. Block graphs are
hereditary; hence, this representative graph, which is $B$, is a
block graph. For a block $P$ of $B$, let $\widehat P$ be the union
of its classes $C_x$ and let $w(P)=|\widehat P|$. Every clique of
$G$ is contained in some $\widehat P$, so $\omega(G)=\max_P w(P)$
when $G$ is nonempty.

Strictly chordal graphs are strongly chordal. Block graphs are chordal
and contain no induced sun, since their maximal cliques meet in at
most one vertex and a sun contains an induced diamond. Adding a true
twin cannot create an induced hole or sun. Neither obstruction has
a pair of true twins, so an obstruction after the addition would
also occur before it. The characterization of strongly chordal
graphs as sun-free chordal graphs gives the inclusion~\cite{Farber1983}.

\begin{theorem}
	\label{thm:strictly-chordal-pcf}
	Let $G$ be a nonempty strictly chordal graph with clique number $k$.
	Then $\chi_{\mathrm{pcf}}(G)=k+1$ if the quotient $B$ has a vertex
	$x$ with $|C_x|=1$ that belongs to two distinct blocks of weight
	$k$. In all other cases, $\chi_{\mathrm{pcf}}(G)=k$. Given adjacency
	lists, Algorithm~\ref{alg:strictly-chordal} constructs an optimal
	coloring in $O(n+m)$ time, where $n=|V(G)|$ and $m=|E(G)|$.
\end{theorem}

\begin{proof}
	Suppose $C_x=\{v\}$ and two blocks $P,Q$ of weight $k$ contain $x$.
	Distinct blocks of $B$ meet in at most one vertex, so $\widehat P$
	and $\widehat Q$ meet only in $v$. In any proper $k$-coloring each
	of the other $k-1$ colors occurs on a neighbor of $v$ in both
	blocks. Thus $v$ has no witness.
	
	Use the palette $[p]$ with $p=k+1$ when this obstruction occurs
	and $p=k$ otherwise. Root each block-cut tree of $B$ at a block
	and color its replacement clique injectively. This includes the
	one-block components, in particular isolated vertices of $B$.
	When coloring a nonroot block $Q$ through its parent cut vertex
	$x$, the already colored vertices of $\widehat Q$ are exactly
	those in $C_x$. The new vertices have no neighbors among the
	earlier vertices outside $C_x$. Indeed, a block containing such
	an edge would connect the subtree at $Q$ to the previously
	processed blocks through a vertex other than $x$, contrary to the
	block-cut tree structure.
	
	Two distinct vertices of a class $C_x$ with $|C_x|\ge2$ witness
	each other in every proper coloring. Every other neighbor of one
	is adjacent to the other and therefore has a different color.
	A vertex in a singleton class at a noncut vertex of $B$ has a clique
	neighborhood and hence has a witness whenever it is nonisolated.
	It remains to supply witnesses to singleton cut vertices.
	
	Consider a cut vertex $x$ with $C_x=\{v\}$, and let $P$ be its
	already colored parent block. At most one block incident with $x$
	has weight $p$. If a child block has weight $p$, use that child
	as the supplying block and choose as witness color any color
	missing from $\widehat P$. Such a color exists because
	$w(P)<p$. The supplying child uses all $p$ colors, so this color
	appears there exactly once. Otherwise let $P$ supply the witness
	and choose the color of any vertex of $\widehat P\setminus\{v\}$.
	In either case, forbid the selected color in each child block
	other than the supplying block. Every child subject to the
	restriction has weight at most $p-1$, so its coloring extends
	injectively while keeping the color of $v$. At a cut vertex with
	$|C_x|\ge2$ the child blocks need no additional restriction.
	
	Each extension preserves properness. At a singleton cut vertex,
	the selected color occurs once among the neighbors in its supplying
	block and is absent from all other incident blocks. Since the
	incident blocks contain all its neighbors, the vertex has a witness.
	The same palette works in every component. The clique lower bound
	and the obstruction prove optimality.
\end{proof}

\begin{algorithm}[t]
	\DontPrintSemicolon
	\KwIn{A strictly chordal graph $G$ in adjacency lists.}
	\KwOut{An optimal proper conflict-free coloring of $G$.}
	\If{$V(G)=\varnothing$}{\KwRet the empty coloring\;}
	Compute the true-twin classes $C_x$, their quotient $B$, the blocks of $B$, and the block-cut forest\;
	Set $w(P)\gets \sum_{x\in V(P)}|C_x|$ for each block $P$ and $k\gets\max_P w(P)$\;
	Set $p\gets k+1$ if a singleton class belongs to two blocks of weight $k$, and set $p\gets k$ otherwise\;
	\ForEach{component of the block-cut forest}{
		Root it at a block and color the root clique injectively from $[p]$\;
		\ForEach{cut vertex $x$ in root-to-leaf order}{
			Let $P$ be the parent block of $x$\;
			\If{$C_x=\{v\}$}{
				\eIf{a child block $S$ of $x$ has weight $p$}{
					Let $a$ be a color absent from $\widehat P$\;
				}{
					Set $S\gets P$ and let $a$ be a color on $\widehat P\setminus\{v\}$\;
				}
			}
			\ForEach{child block $Q$ of $x$}{
				Preserve the colors on $C_x$ and color $\widehat Q\setminus C_x$ injectively
				from the remaining colors in $[p]$, excluding $a$ if $|C_x|=1$ and $Q\ne S$\;
			}
		}
	}
	\KwRet the resulting coloring\;
	\caption{Optimal proper conflict-free coloring of a strictly chordal graph.}
	\label{alg:strictly-chordal}
\end{algorithm}

\subsection{Running time}
To compute the classes $C_x$, start with a single part and refine
the partition by each closed neighborhood $N_G[v]$. For a
refinement, group its vertices by their current parts and split
each touched part into vertices in $N_G[v]$ and vertices outside
it. With linked membership lists, part pointers, and temporary
lists of marked vertices, each refinement takes $O(|N_G[v]|)$ time.
The total is $O(n+m)$. Two vertices are in the same final part
exactly when they belong to the same closed neighborhoods, which
by symmetry is equivalent to having equal closed neighborhoods.
Scan the neighbors of one representative per class to build the
quotient, discarding duplicate class identifiers with marks. Find
its blocks and the block-cut forest by depth-first
search~\cite{Tarjan1972}. Block weights and the number of incident
blocks of weight $k$ at each cut vertex are computed by scanning
the block incidences.

To extend a block of weight $w(P)$, mark the colors already assigned
to its parent class and the one forbidden color, if present. Scan
the palette in order until all new vertices are colored. Before
the scan stops, each inspected color is assigned or is one of
the marks. At most $w(P)+1$ colors are inspected. After each block
is colored, cache the colors of any two distinct vertices if
there are two, and one missing color if the block is smaller than
the palette. The missing color can be found by marking the colors
of the block and inspecting at most $w(P)+1$ palette entries.
These cached values give either kind of witness color at a
singleton cut vertex in constant time. Timestamps let us reuse
the mark array across blocks.

It remains to bound the total weight of the blocks. For a
nonisolated vertex $x$ of $B$, the number of blocks containing $x$
is at most $\deg_B(x)$, since each such block contains a different
edge incident with $x$. Thus $\sum_P w(P)$ is at most
$n+\sum_{xy\in E(B)}(|C_x|+|C_y|)$. Each quotient edge $xy$ accounts
for $|C_x||C_y|$ edges of $G$, and
$|C_x|+|C_y|\le 2|C_x||C_y|$. Hence $\sum_P w(P)\le n+2m$.
All the scans and coloring steps therefore take $O(n+m)$ time.

When all replacement cliques are singletons, Theorem~\ref{thm:strictly-chordal-pcf}
recovers the block-graph characterization of Pradhan and
Sharma~\cite{PradhanSharma2026}. A separator of two or more true twins
needs no witness from either neighboring block, even when both blocks
have weight $k$. This yields tractable instances with unbounded clique
separators. For example, subdivide every edge of a claw and replace one
subdivision vertex by a clique of arbitrary size. The resulting
strictly chordal graph contains the subdivided claw as an induced
subgraph. In an interval model of the subdivided claw, the three
subdivision intervals would be disjoint and meet the center interval.
The middle one would lie entirely inside the center interval, so its
terminal neighbor could meet it only by meeting the center. Hence these
examples lie outside the interval class.

\section{Strongly chordal graphs with small clique-tree separators}
\label{sec:strong-small-separators}

For a chordal graph, a clique tree has one node for each maximal clique, and the nodes containing any fixed vertex induce a connected subtree. At an edge of a clique tree, the intersection of the two incident cliques separates the vertices on its two sides. We use the maximum size $s$ of these intersections in a supplied clique tree, taking $s=0$ for a one-node tree. The clique number is unrestricted.

\begin{theorem}
	\label{thm:strong-small-separators}
	Let $G$ be a strongly chordal graph on $n$ vertices, and suppose that a clique tree of $G$ is supplied in which every separator has size at most $s$. Given $k$, one can decide whether $G$ has a proper conflict-free $k$-coloring and construct such a coloring when one exists in $n^{2^{O(s+1)}}$ time. Consequently, for every fixed $s$, an optimal proper conflict-free coloring can be found in polynomial time.
\end{theorem}

We may work with connected components separately. If $k\ge n$, assign distinct colors to all vertices. If $k<\omega(G)$, reject by properness. Thus we may assume $\omega(G)\le k\le n$ and use the palette $[k]=\{1,\dots,k\}$. Counts are always truncated by $\tau(t)=\min\{t,2\}$. This truncation commutes with addition in the sense that $\tau(a+b)=\tau(\tau(a)+\tau(b))$.

\subsection{The structure of the child separators}

Root the clique tree. Fix a clique node $Q$ and let $Q_1,\dots,Q_d$ be its children. Put $H_i=Q\cap Q_i$. For every child, some vertex $u_i\in Q_i\setminus Q$ has exactly $H_i$ as its neighborhood in $Q$. To see this, any clique containing $u_i$ is in the child subtree. If a vertex of $Q$ is adjacent to $u_i$, there is a clique containing both. By running intersection that vertex is also in $Q_i$, and hence in $H_i$. Vertices $u_i$ from distinct child subtrees are nonadjacent.

\begin{lemma}
	\label{lem:child-separators-acyclic}
	The incidence graph of the sets $H_1,\dots,H_d$ on the ground set $Q$ has no induced cycle of length at least six. The graph on $Q$ obtained by making every $H_i$ a clique has a tree decomposition with bags of size at most $\max\{s,1\}$, in which every $H_i$ is contained in a bag. Such a decomposition can be computed in polynomial time.
\end{lemma}

\begin{proof}
	An induced incidence cycle of length $2t$ with $t\ge 3$ would select $t$ distinct vertices of $Q$ and $t$ child separators. Together with the corresponding independent vertices $u_i$, the selected vertices of $Q$ would induce a $t$-sun. Strong chordality forbids this~\cite{Farber1983}.
	
	We describe how to construct the small bags. A nest point in a set family is a vertex whose incident sets are linearly ordered by inclusion. At every restriction to a subset $U\subseteq Q$, the restricted child separator family has a nest point if $U$ is nonempty. For a direct proof, form a split graph with one vertex for each restricted set on its clique side and the vertices of $U$ on its independent side. Use membership as adjacency between the two sides. An induced sun would have its central clique entirely on the clique side, since an independent-side vertex has a clique neighborhood. Each ray must then lie on the independent side, since a clique-side vertex would see all the central vertices. The sun would therefore give an induced incidence cycle. The split graph is chordal and sun-free, hence strongly chordal. By Farber's simple-vertex characterization~\cite{Farber1983}, it has a vertex whose closed neighborhood consists of vertices with pairwise comparable closed neighborhoods. If that vertex belongs to the independent side, the set vertices adjacent to it have nested neighborhoods in $U$, so it is a nest point. If it belongs to the clique side, all the set vertices have nested neighborhoods in $U$, in which case any vertex of $U$ is a nest point.
	
	Repeatedly delete a nest point. In the graph formed by making each $H_i$ a clique, the later neighbors of the deleted vertex are precisely the other vertices of its largest incident restricted set, or are empty. They form a clique and, when nonempty, have size at most $s-1$. For each deleted vertex, make a bag containing it and its later neighbors. If the later neighbors are nonempty, connect its bag to the bag of the earliest deleted vertex among them. This latter bag contains all its other later neighbors. Connect bags with no parent through an empty bag. Each edge is in the bag of its earlier endpoint, and the bags containing any vertex are connected by the parent links. These bags therefore form a tree decomposition. Each nonempty $H_i$ lies in the bag of its earliest deleted member. Testing whether a point is a nest point takes polynomial time, so the entire construction is polynomial.
\end{proof}

\subsection{What a subtree communicates to its parent}

For a clique node $Q$, let $V_Q$ be the union of the cliques in its rooted subtree and let $S$ be its intersection with its parent clique, taking $S=\varnothing$ at the root. The vertices in $V_Q\setminus S$ have no neighbors outside $V_Q$. Our table records the proper colorings of $G[V_Q]$ in which every nonisolated vertex of $V_Q\setminus S$ already has a witness. Vertices of $S$ are left unchecked.

For every color $a\in[k]$, record its \emph{count vector} on $S$. At coordinate $v\in S$ the entry is $\tau$ of the number of neighbors of $v$ in $V_Q\setminus S$ colored $a$. The distinct colors assigned to vertices of $S$ are distinguished, one for each vertex of $S$. For the other colors, it is enough to record how many have each of the $3^{|S|}$ possible vectors. This is valid because a common permutation of all colors that fixes the distinguished colors preserves every requirement in the subtree. We call these data a \emph{signature}. A realizable signature is the information passed to the parent. In particular, the signature table at the root is nonempty exactly when a proper conflict-free $k$-coloring exists.

We next show how to compute the table at $Q$ from the child tables without enumerating all color assignments to the potentially large clique $Q$.

\subsection{Combining the child signatures at a clique}

Fix a palette pool $P\subseteq[k]$ of size $|Q|$ for the clique $Q$. There is no loss in fixing this pool, since we can permute colors in $G[V_Q]$ before passing the signature to the parent. Fix distinct members of $P$ as the colors of the vertices of $S$. At the end of the computation every color of $P$ will occur on exactly one vertex of $Q$.

Use the tree decomposition from Lemma~\ref{lem:child-separators-acyclic} and add $S$ to every bag. Its bags now have at most $s+\max\{s,1\}$ vertices. Root it, and convert it to a nice tree decomposition with leaf and root bags equal to $S$. This can be done by replacing each tree edge by a sequence of single-vertex additions and deletions and using binary join nodes with equal bags. For each child $Q_i$, assign its signature table to a bag containing $H_i$, and insert a copy of that bag as a factor node. The purpose of that node is to incorporate exactly one signature from the child table.

The following states are used in this local tree decomposition. Its current bag is denoted by $X$, and $S\subseteq X$ throughout.
\begin{enumerate}
	\item For every $v\in X$, store its color $c(v)\in P$. These colors are distinct. For each nonisolated $v\in X\setminus S$, also store a proposed witness color $w(v)\ne c(v)$ and the capped number of occurrences of $w(v)$ among its neighbors contributed by child factors processed so far. This count excludes vertices of $Q$.
	\item Distinguish every color named by the data in item~1 or by the fixed colors on $S$. For each distinguished color, store its pool membership, whether it has already been assigned to a forgotten vertex of $Q$, and its capped count vector on $S$ from processed child factors.
	\item For all other colors, store a histogram indexed by pool membership, assignment to a forgotten vertex of $Q$, and the count vector on $S$ from processed child factors.
\end{enumerate}

A state identifies named colors only by their equality pattern and their recorded attributes. In particular, the histogram entries count unnamed colors, rather than retaining their actual labels. The color of an active vertex cannot already be assigned to a forgotten vertex. A proposed witness color may be the color of another active vertex, a forgotten vertex, or a vertex of $Q$ not yet introduced.

In a realization of a state, the processed part of the local decomposition has an injective assignment of colors from $P$ to its introduced vertices of $Q$, and has chosen realizable signatures for the child factors in that part. Each child coloring agrees with the assignment on its separator. The state records exactly the contributions of these child colorings to its witness counters and to the parent count vectors. Every forgotten nonisolated vertex of $Q\setminus S$ has passed its witness test, counting one occurrence inside $Q$ whenever its witness color belongs to $P$. The completed coloring will use the pool bijectively on $Q$, so any color of $P$ different from the vertex's own color will occur on exactly one of its neighbors in $Q$.

\subsection{Leaf and introduce nodes}
At a leaf, only $S$ is present, no colors have been assigned to forgotten vertices, and all counts from child factors are zero. At an introduce node for $v\notin S$, choose a color of $P$ not used by any active or forgotten vertex. If $v$ is nonisolated, choose a witness color different from $c(v)$ and initialize its child count to zero. A chosen color may be a distinguished color or may come from a nonempty histogram category. In the latter case, remove one member from the category and give it a distinguished name. The initial witness count is zero because no processed child factor could contain $v$ in its separator before $v$ is present in the local decomposition.

\subsection{Child factor nodes}
At the factor for $Q_i$, choose one of its realizable signatures. Map its distinguished colors on $H_i$ to the colors assigned to the vertices of $H_i$ in $X$. For its unnamed colors, enumerate all possible alignments with the remaining local colors. This is done by a contingency table. A row specifies a count-vector category from the child signature. A column specifies either one distinguished local color not already mapped or an unnamed local category. The row and column sums must agree with the numbers of colors of each kind. Such a table represents a bijection because colors within a category can be permuted arbitrarily. The fixed alignment of colors on $H_i$ ensures that the renamed child coloring agrees with $c$ there.

For $v\in H_i\setminus S$, add the contribution for $w(v)$ to its witness counter. For each parent vertex $p\in S\cap H_i$, add the appropriate contribution to coordinate $p$ of every color's count vector. All additions are capped at two. The other coordinates are unchanged, since a vertex of $Q\setminus H_i$ has no neighbors in the exclusive part of the child subtree. The contingency table gives the updated histogram by grouping the colors by their new attributes. Colors used in the child outside $Q$ do not change the record of which colors have been assigned to vertices of $Q$.

\subsection{Forget nodes}
When forgetting a nonisolated vertex $v\notin S$, require its child count at color $w(v)$ to be zero if $w(v)\in P$, and one otherwise. In the former case its unique occurrence is supplied by another vertex of $Q$. In the latter case the witness lies in a child subtree. All child factors involving $v$ have been processed when this check is made. Mark $c(v)$ as assigned to a forgotten vertex. Any color no longer named by the active data can be returned to the histogram with its current attributes. An isolated vertex needs no witness test.

\subsection{Join nodes}
The two children have the same active bag. Their equality patterns for the colors and witnesses of the active vertices must agree, along with the prescribed colors on $S$. Align these distinguished colors accordingly and require their pool membership to agree. For the unnamed colors, enumerate contingency tables matching categories from the two children with equal pool membership. Reject any match in which a color has been assigned to a forgotten vertex on both sides. Apply the same test to the distinguished colors. The forgotten vertices in the two subtrees are disjoint, so this is precisely the remaining injectivity constraint for $Q$. The assignment flag of a merged color is the union of its two flags. Add the count vectors and active witness counters with truncation. Every compatible contingency table can be realized by permuting the unnamed colors in one child, and that permutation preserves all tests already passed.

At the root of the local decomposition, the active bag is $S$. All vertices of $Q\setminus S$ have passed their witness tests. The use of distinct colors from a pool of size $|Q|$ ensures that each pool color is assigned to one vertex of $Q$. Add one to every coordinate of the parent count vector of each color assigned to $Q\setminus S$, since every such vertex is adjacent to every vertex of $S$. Discard the pool and assignment attributes and retain only the distinguished count vectors and the histogram of the other vectors. This is the signature passed from $Q$ to its parent.

\begin{algorithm}[t]
	\DontPrintSemicolon
	\KwIn{A strongly chordal graph, a clique tree with separator size at most $s$, and a palette of $k$ colors.}
	\KwOut{A proper conflict-free coloring with at most $k$ colors, if one exists.}
	Root the clique tree and process its cliques in bottom-up order\;
	\ForEach{clique $Q$}{
		Form the local tree decomposition of the child separators and add the parent separator $S$ to its bags\;
		Fix a pool of $|Q|$ colors and the colors on $S$\;
		Compute local states at introduce, child factor, forget, and join nodes\;
		Add the contributions from $Q\setminus S$ and record all realizable signatures at $S$\;
	}
	Use a realizing state at the root to recover the coloring by backtracking, or report that no coloring exists if there is no such state\;
	\caption{Exact PCF coloring for bounded clique-tree separator size.}
	\label{alg:small-separators-pcf}
\end{algorithm}

\subsection{Correctness and running time}

The transition invariant follows by induction over the local decomposition. The introduce and forget transitions implement precisely the clique-coloring and witness requirements. At a child factor, every feasible child coloring induces one of the enumerated alignments, and each enumerated alignment is realized by a permutation of a child coloring. At a join, after matching the distinguished colors and the unnamed categories, the two partial colorings agree on the common active vertices. No forgotten vertices of $Q$ receive the same color across the two pieces. All contributions from distinct child subtrees add, and those contributions include no vertex of $Q$. Thus neither a neighbor nor a color assignment is counted twice. Conversely, any coloring realizing a state induces one of the matching tables and a valid state on each side. Since the clique pool is used bijectively at completion, the promised within-clique contribution in every forget test is correct. The table of signatures at $Q$ is therefore exact. Induction up the original clique tree proves correctness of Algorithm~\ref{alg:small-separators-pcf}. Store the choices for each state, including the contingency tables. Assigning concrete names within each category during backtracking recovers the coloring.

There are $O(s+1)$ distinguished labels at any local node. The number of unnamed categories is at most $4\cdot 3^s$, from the two pool choices, two assignment flags, and the count vectors. Each histogram entry is at most $k\le n$. The equality patterns, witness counters, and distinguished attributes contribute a factor depending only on $s$, bounded by $2^{2^{O(s+1)}}$. A contingency table has $2^{O(s+1)}$ entries, each at most $n$. Consequently, the number of states and the time to enumerate transitions at a node are bounded by $n^{2^{O(s+1)}}$. There are polynomially many cliques, local bags, and factor nodes, giving the time bound in Theorem~\ref{thm:strong-small-separators}. To find an optimum, test the palette of size $\omega(G)$. If it is infeasible, Theorem~\ref{thm:sc-additive} supplies an optimal coloring with one more color.

If a clique tree is not supplied, the existence of one with separators of size at most $s$ can also be tested in polynomial time. On the maximal cliques, give a potential tree edge $QQ'$ weight $|Q\cap Q'|$ and retain only edges of weight at most $s$. A spanning tree has total weight at most $\sum_{v\in V(G)}(m_v-1)$, where $m_v$ is the number of maximal cliques containing $v$. This follows by counting, for each $v$, the tree edges whose endpoints both contain $v$. Equality holds exactly when those cliques induce a connected subtree for every $v$, which is the clique-tree condition. If the retained edges have no spanning tree, then no clique tree with the required separator bound exists. Otherwise, compute a maximum-weight spanning tree and accept it exactly when its weight reaches the bound. This argument can be applied separately in each connected component.

\section{Concluding questions}
\label{sec:conclusion}

A star already suffices as the host tree for the path-graph
hardness result. The constructed graphs can contain an induced
sun, which places them outside the strongly chordal class.
For strongly chordal graphs, the additive-one bound in
Theorem~\ref{thm:sc-additive} reduces optimization to deciding
whether the clique number of colors suffices. We resolve this
decision under several conditions on the maximum cliques, for
strictly chordal graphs, and for bounded clique-tree separators.
The complexity on all strongly chordal graphs, and even on
all interval graphs, remains open. It is also open whether
the dependence on the separator size can be moved out of the
exponent of $n$ in Theorem~\ref{thm:strong-small-separators}.

For chordal bipartite graphs, the hardness proof applies to
every fixed palette of size at least four. The complexity of
three-colorability under the proper conflict-free requirement
remains open on this class. Proposition~\ref{prop:chordal-bipartite-small-palettes}
excludes a three-coloring on nonempty instances of minimum
degree at least two. The remaining cases allow leaves, which can
supply witnesses to their neighbors.

\bibliographystyle{plain}
\bibliography{PCF_ref}
         
\end{document}